\documentclass[12pt]{article}
\usepackage[cp1250]{inputenc}
\usepackage[verbose,a4paper,tmargin=0.5in,lmargin=1in,rmargin=1in]{geometry}
\usepackage[pdftex]{graphicx}
\usepackage{amsfonts}
\usepackage{indentfirst}
\usepackage{latexsym}
\usepackage{amsmath}
\usepackage{amsthm}
\usepackage{amssymb}
\usepackage{psfrag}
\usepackage[mathscr]{eucal} 
\usepackage{color}
\usepackage{cite}
\usepackage{verbatim}

\title{Classification of complex and real, vacuum spaces of the type $[\textrm{N}] \otimes [\textrm{N}]$.}

\author{$\textrm{Adam Chudecki}^{*}$}

\begin{document}

\maketitle

$*$ Center of Mathematics and Physics, Lodz University of Technology, 
\newline
$\ \ \ \ \ $ Al. Politechniki 11, 90-924 Łódź, Poland, adam.chudecki@p.lodz.pl
\newline
\newline
\newline
\textbf{Abstract}. 
Complex and real, vacuum spaces with both self-dual and anti-self-dual parts of the Weyl tensor being of the type [N] are considered. Such spaces are classified according to two criteria. The first one takes into account the properties of the congruences of totally null, geodesic 2-dimensional surfaces (the null strings). The second criterion is based on investigations of the properties of the intersection of these congruences. It is proved that there exist six distinct types of the  $[\textrm{N}] \otimes [\textrm{N}]$ spaces. New examples of the Lorentzian slices of the complex metrics are presented. Also, some type $[\textrm{N}] \otimes [\textrm{N}]$ spaces which do not posses Lorentzian slices are considered.

\section{Introduction}

In 1976 J.F. Plebański and I. Robinson introduced the concept of \textsl{hyperheavenly spaces} ($\mathcal{HH}$-spaces) \cite{Plebanski_Robinson_1,Plebanski_Robinson}. $\mathcal{HH}$-spaces are defined as a 4-dimensional complex manifolds which satisfy Einstein vacuum field equations and which self-dual (SD) or anti-self-dual (ASD) part of the Weyl tensor is algebraically degenerate. It appeared that in $\mathcal{HH}$-spaces Einstein vacuum field equations could be reduced to the single, partial differential equation of the second order (\textsl{hyperheavenly equation}) for one holomorphic function of four variables called \textsl{the key function}. It seemed that $\mathcal{HH}$-spaces offer a new technique of finding exact solutions of Einstein field equations in vacuum. This technique consists of three steps. First step: specialize hyperheavenly equation for such $\mathcal{HH}$-spaces, that SD and ASD parts of the Weyl tensor are of the same Petrov - Penrose type. Second step: solve the hyperheavenly equation (possibly with symmetries). Third step: find real Lorentzian slice of the complex metric generated by the solution of the hyperheavenly equation. Particularly, $\mathcal{HH}$-spaces give a new hope in finding twisting vacuum type [N] solutions. As is known, all vacuum Lorentzian nontwisting type [N] metrics have been found (pp-waves, Kundt class) or reduced to the single PDE of the second order (Robinson - Trautman class). In contrary to the nontwisting case, only one vacuum Lorentzian twisting type [N] metric is explicitly known. This is the famous Hauser metric found in 1974 \cite{Hauser}. A great effort has been done to find other metrics but without a considerable success. It has been expected that it is only a matter of time when some new twisting type [N] solutions will be found as a Lorentzian slices of the type $[\textrm{N}] \otimes [\textrm{N}]$ $\mathcal{HH}$-spaces.  

The first step of the technique described above has been fulfilled. In \cite{Plebanski_Torres} the key function for the type $[\textrm{N}] \otimes [\textrm{N}]$ spaces has been presented in the useful form (see also \cite{Chudecki_Przanowski} for slightly different approach). Unfortunately, hyperheavenly equation for the complex vacuum type $[\textrm{N}] \otimes [\textrm{N}]$ space still appears to be very hard to solve. Hence, the second step is a serious challenge. There are a few attempts which investigate the vacuum twisting type $[\textrm{N}] \otimes [\textrm{N}]$ spaces equipped with two symmetries \cite{Finley_1, Chudecki_Przanowski}, but the result is always the same: extremely advanced ODE, much more complicated then the equation which appears in Hauser solution. $\mathcal{HH}$-spaces of the type $[\textrm{N}] \otimes [\textrm{N}]$ with one symmetry or without any symmetries are even harder to solve \cite{Finley_2,Finley_Plebanski,Chudecki_Przanowski}. Hyperheavenly equation splits into the system of equations which is overdetermined. To make matters worse, the third step is even more complicated that anyone could suspect. Except general analysis \cite{Rozga} and a few examples \cite{Boyer_Finley_Plebanski,Chudecki_klasyfikacja_Killingow_HH_nieeks}, no general techniques of obtaining real Lorentzian slices have been developed so far. 

It does not mean, of course, that $\mathcal{HH}$-spaces are another dead end in the theoretical physics. They play a great role in the 4-dimensional neutral geometries. Indeed, although it is very hard to obtain real Lorentzian spaces from the $\mathcal{HH}$-spaces, real spaces equipped with the metric of the neutral signature $(++--)$ can be obtained very easily. It is enough to replace all complex coordinates by the real ones and all holomorphic functions by the real analytic ones. Recently 4-dimensional neutral geometries find their applications in many issues of theoretical physics (Walker manifolds \cite{Garcia_Rio}, Osserman manifolds \cite{Garcia_Rio_2}, integrable systems and ASD structures \cite{Dunajski_Tod}, rolling bodies \cite{Nurowski_1, Nurowski_2}).

The present paper is devoted to the complex and real vacuum $\mathcal{HH}$-spaces of the type $[\textrm{N}] \otimes [\textrm{N}]$ and it has three main purposes. Namely we are going to:
\begin{enumerate}
\item obtain new examples of Lorentzian slices of the complex metrics.
\newline
Nontwisting, vacuum, real Lorentzian type [N] spaces are all known. It gives the opportunity to investigate the question what steps should be taken to reconstruct those solutions from the generic complex solutions. We succeeded in reconstruction the Kundt class and the Robinson - Trautman solution. We believe, that these new examples of Lorentzian slices of complex metrics are another step toward more ambitious task, namely: general techniques of obtaining Lorentzian slices.
\item describe type $[\textrm{N}] \otimes [\textrm{N}]$ spaces which do not admit any real Lorentzian slice.
\newline
$\mathcal{HH}$-spaces of the type $[\textrm{N}] \otimes [\textrm{N}]$ are equipped with congruences of SD and ASD totally null and totally geodesic 2-dimensional surfaces, called \textsl{the null strings}. The important property of such family of surfaces is \textsl{expansion} (see subsection \ref{subsekcja_intersection} for details). Spaces equipped with expanding congruence of SD null strings and nonexpanding congruence of ASD null strings (or vice versa) are interesting examples of the spaces which do not admit any Lorentzian slice. There are two subtypes of $[\textrm{N}] \otimes [\textrm{N}]$ spaces with such a property. Considered as a real neutral spaces they appear to be the Walker spaces. We investigate this in detail in the present paper.
\item find the detailed classification of the $[\textrm{N}] \otimes [\textrm{N}]$ spaces.
\newline
Usually $\mathcal{HH}$-spaces are classified according to properties of the congruences of the null strings. We consider another criterion: the properties of the intersection of these congruences.
\end{enumerate}

In what follows it is assumed that cosmological constant vanishes. Hence, only vacuum case is considered. All considerations are purely local and complex (functions are holomorphic, coordinates are complex). If coordinates are considered as the real ones and functions as the real analytic ones then the corresponding space is real with neutral signature metric. Consequently, all the metrics presented in the paper have real neutral slices. To obtain real Lorentzian slices some subtle steps have to be introduced.

The paper is organized, as follows.

Section \ref{Przestrzenie_Nxany_preliminaries} is devoted to reminding basic knowledge on expanding hyperheavenly spaces (definition, the metric, coordinate gauge freedom, symmetries). Also some basic information about the congruences of SD and ASD null strings is presented. In section \ref{section_przestrzenie_NxN} we focus on the complex and real type $[\textrm{N}] \otimes [\textrm{N}]$ spaces. Properties of the intersection of the congruences of SD and ASD null strings are investigated and new approach to the problem of the classification of the type $[\textrm{N}] \otimes [\textrm{N}]$ spaces is presented. There are six different subtypes of the vacuum type $[\textrm{N}] \otimes [\textrm{N}]$ spaces. Also, the key function for every type of the considered spaces is found.

The main aim of the rest of the paper (sections \ref{Sekcja_pp_fale}-\ref{Sekcja_klasa_III_szczegolna}) is to consider the corresponding types in details. Einstein field equations in vacuum are completely solved or reduced to the single equation. Much attention has been devoted to the symmetries which are investigated in two steps. Firstly, we equip the space with single homothetic vector, i.e., the vector which satisfies the equations $\nabla_{(a} K_{b)} = \chi_{0} g_{ab}$. The second step consists of specializing the results for the $\chi_{0}=0$ (homothetic vector is reduced to the Killing vector). Then the second homothetic vector is admitted. Such approach covers all possible cases with one symmetry (one homothetic or one Killing vector) and with two symmetries (one Killing vector, one homothetic vector or two Killing vectors). Also, new examples of the Lorentzian slices of the complex metrics are presented (subsections \ref{Subsekcja_klasa_I_bez_symetrii}, \ref{Subsubsekcja_klasa_II_Bjeden_bez_symetrii} and \ref{Subsubsekcja_klasa_II_Bzero_bez_symetrii}). Finally, the spaces considered in the sections \ref{Sekcja_klasa_I_b} and \ref{Sekcja_klasa_III_szczegolna} admit only real neutral slices. Only very special classes of these spaces were considered earlier \cite{Chudecki_null_Killing,Chudecki_Przanowski}.


\setcounter{equation}{0}
\section{Expanding $\mathcal{HH}$-spaces of the type $[\textrm{N}] \otimes [\textrm{any}]$}
\label{Przestrzenie_Nxany_preliminaries}

\subsection{The metric}

In this section we present the brief summary of the formalism of \textsl{expanding hyperheavenly spaces ($\mathcal{HH}$-spaces)} (see, e.g., \cite{Plebanski_Robinson,Finley_Plebanski_1}, see also our previous paper \cite{Chudecki_Przanowski}). Except expanding $\mathcal{HH}$-spaces there are also \textsl{nonexpanding $\mathcal{HH}$-spaces}. The difference between them will be pointed out in the subsection \ref{subsekcja_kongruencje_strun_SD}. For our purposes it is necessary to collect the basic facts about the first type of $\mathcal{HH}$-spaces, i.e. about the expanding ones.  

$\mathcal{HH}$-space is a 4-dimensional complex analytic differential manifold equipped with the holomorphic metric $ds^{2}$ such that the self-dual (SD) or anti-self-dual (ASD) part of the Weyl tensor is algebraically degenerate and such that the Einstein vacuum field equations are satisfied $R_{ab}=0$. 

In what follows we choose the orientation in such a manner that SD part of the Weyl tensor is algebraically degenerate and moreover it is of the type [N]. Consequently, we deal with the spaces of the type $[\textrm{N}] \otimes [\textrm{any}]$. The metric of such spaces reads
\begin{eqnarray}
\label{metryka_typu_Nxany}
ds^{2} &=& 2\phi^{-2} \big\{ \tau^{-1} (d \eta  d w - d \phi  dt) -    \phi \, W_{\eta\eta} \,   
dt^{2} \ \ \ \ \ \ 
\\ \nonumber
&& \ \ \ \ \ \ \ \ \ \ + ( 2W_{\eta} - 2\phi \, W_{\eta \phi}  ) \, dw dt
  +  ( 2  W_{\phi}  - \phi \, W_{\phi \phi}  ) \, dw^{2} \big\} \ \ \ \ \ \ 
\end{eqnarray}
where $(\phi, \eta, w,t)$ are local coordinates called \textsl{Plebański - Robinson - Finley (PRF) coordinates} and $\tau$ is nonzero arbitrary constant. We use abbreviations $W_{\eta} := \partial W / \partial \eta$, $W_{\eta \phi} := \partial^{2} W / \partial \eta \partial \phi$, etc.. Function $W=W(\phi, \eta, w,t)$ is called \textsl{the key function}. Vacuum Einstein field equations are reduced to a single, nonlinear PDE of the second order, which is called \textsl{the expanding hyperheavenly equation}
\begin{equation}
\label{rownanie_HH}
   W_{\eta \eta}W_{\phi \phi} - W_{\eta \phi}^{2} + 2 \phi^{-1} W_{\eta} W_{\eta \phi} - 2 \phi^{-1} W_{\phi}W_{\eta\eta}  + (\tau \phi)^{-1} ( W_{w\eta}-W_{t\phi} )
 = \frac{ \gamma}{\tau^{2}}
\end{equation}
where $\gamma = \gamma(w,t)$ is an arbitrary function called \textsl{the structural function}. Spinorial images of the SD and ASD conformal curvature read
\begin{subequations}
\begin{eqnarray}
&& C_{1111} = C_{1112} = C_{1122} = C_{1222} = 0 , \ C_{2222} = \tau \phi^{7} \gamma_{t} 
\\ 
\label{ASD_conformal_curvature_ogolnie}
&& C_{\dot{A}\dot{B}\dot{C}\dot{D}} = \phi^{3} \, W_{p^{\dot{A}}p^{\dot{B}}p^{\dot{C}}p^{\dot{D}}}
\end{eqnarray}
\end{subequations}
SD conformal curvature of the type [N] is characterized by the condition $\gamma_{t} \ne 0$, otherwise the space reduces to the right conformally flat space. In (\ref{ASD_conformal_curvature_ogolnie}) "spinorial notation" of the coordinates $(\phi,\eta,w,t)$ is used. Relation between $(\phi,\eta,w,t)$ and $(p^{\dot{A}}, q^{\dot{B}})$ is given by the equations
\begin{equation}
\label{zwiazaki_miedzy_wspolrzednymi}
\tau p^{\dot{A}} = \eta J^{\dot{A}} + \phi K^{\dot{A}} \ , \ \ \ \tau q^{\dot{A}} = t J^{\dot{A}} + w K^{\dot{A}}
\end{equation}
where $J_{\dot{A}}$ and $K_{\dot{A}}$ are constant spinors such that 
\begin{equation}
K^{\dot{A}} J_{\dot{A}} = \tau \ne 0
\end{equation}
Geometrical interpretation of the PRF-coordinates is explained in the subsection \ref{subsekcja_kongruencje_strun_SD}.

\subsection{Coordinate gauge freedom}

$\mathcal{HH}$-spaces of the type $[\textrm{N}] \otimes [\textrm{any}]$ with $\Lambda = 0$ admit the coordinate gauge freedom. The metric (\ref{metryka_typu_Nxany}) remains invariant under coordinate transformations

\begin{equation}
\label{coordinate_transformations}
w'=w'(w), \ t'=t'(t,w), \ \phi' = \phi \, \lambda^{-\frac{1}{2}} , \ \eta' = \frac{\lambda^{-1}}{w'_{w}} \, \eta +
     \lambda^{-\frac{1}{2}} \frac{t'_{w}}{w'_{w}} \, \phi + \tau \sigma
\end{equation}
where $\sigma=\sigma(w,t)$ is an arbitrary function and $\lambda=\lambda(w,t)$ is the function such that $t'_{t} =: \lambda^{-\frac{1}{2}}$. [Note that the coordinate $w$ is somehow distinguished; this interesting fact has a deep geometrical interpretation and it will be explained in the subsection \ref{subsekcja_kongruencje_strun_SD}]. The coordinate transformations (\ref{coordinate_transformations}) and the additional gauge freedom of $W$ give
\begin{eqnarray}
\label{transformacja_funkcji_kluczowej}
&& (w'_{w})^{2} \lambda^{\frac{3}{2}} W' = W - \frac{1}{3} L \, \phi^{3}-\frac{1}{2} \lambda w'_{w} ( \sigma_{t} \eta + \sigma_{w} \phi ) - M
\\ \nonumber
&& \ \ \ \ \ \ \ \ \ \ \ \  - \frac{1}{2\tau} \lambda^{\frac{1}{2}} w'_{w} 
\left[   \frac{\partial}{\partial t} \left( \frac{t'_{t}}{w'_{w}} \right) \eta^{2}
+  \left( \frac{\partial}{\partial t} \left( \frac{t'_{w}}{w'_{w}} \right) +  \frac{\partial}{\partial w} \left( \frac{t'_{t}}{w'_{w}} \right) \right) \eta \phi
+   \frac{\partial}{\partial w} \left( \frac{t'_{w}}{w'_{w}} \right) \phi^{2} \right]
\end{eqnarray}
and
\begin{equation}
(w'_{w})^{2} \gamma' = \gamma  - (w'_{w})^{\frac{1}{2}} [(w'_{w})^{-\frac{1}{2}}]_{ww} 
\end{equation}
where $L=L(w)$ and $M=M(w,t)$ are arbitrary gauge functions.

\subsection{Congruences of the SD null strings}
\label{subsekcja_kongruencje_strun_SD}

Complex Goldberg - Sachs theorem \cite{Przanowski_Plebanski_twSachsa} says that any $\mathcal{HH}$-space such that SD part of Weyl tensor is algebraically degenerate admits existence of \textsl{congruences of SD null strings}. It means that for every point $p$ of $\mathcal{HH}$-space there exist an open neighborhood $U$ of $p$ and a 2-dimensional complex totally null SD distribution $\mathcal{D}_2$ on $U$ which appears to be completely integrable. Family of integral manifolds of the distribution $\mathcal{D}_2$ constitute the congruence (also called \textsl{foliation}) of totally null, totally geodesic, SD complex 2-dimensional surfaces called \textsl{SD null strings}. The number of distinct congruences of SD null strings is equal to the number of independent Penrose spinors. Because the space of  type $[\textrm{N}] \otimes [\textrm{any}]$ has only one undotted Penrose spinor, there is only one congruence of SD null strings and it is uniquely determined by the Penrose spinor. 

If we denote SD Penrose spinor by $a_{A}$ then $C_{ABCD} =C a_{A} a_{B} a_{C} a_{D}$ (where $C$ is proportionality factor which can be absorbed into spinor $a_{A}$, if desired) and the following \textsl{SD null string equations} are satisfied
\begin{equation}
\label{rownania_strun_SD}
  a^{B} \nabla_{A \dot{C}} a_{B} = a_{A} M_{\dot{C}}
\end{equation}
The crucial property of the congruence of SD null strings is burried in 1-index dotted spinor field $M_{\dot{C}}$ which is called \textsl{the expansion of the congruence of SD null strings}. If $M_{\dot{C}}=0$ then such congruence is called \textsl{nonexpanding}, if $M_{\dot{C}} \ne 0$ then it is \textsl{expanding}. Nonexpanding congruence is parallely propagated, i.e.
\begin{eqnarray}
\nonumber
M_{\dot{A}}=0  &\Longleftrightarrow &  \nabla_{X}V \in \mathcal{D}_{2} \textrm{ for every vector field } V \in \mathcal{D}_{2} 
\\ \nonumber
&& \textrm{ and for arbitrary vector field } X
\end{eqnarray}
Consequently, if $\mathcal{HH}$-space is equipped with the expanding congruence of SD null strings we call such a space \textsl{expanding $\mathcal{HH}$-space}. \textsl{Nonexpanding $\mathcal{HH}$-spaces} are equipped with the nonexpanding congruences of SD null strings. 

Eqs. (\ref{rownania_strun_SD}) in expanding $\mathcal{HH}$-spaces can be rewritten in the form 
\begin{equation}
\label{rownania_strun_SD_rozpisane}
a_{A} M_{\dot{C}} = a^{B} \partial_{A \dot{C}} a_{B} + a^{B} a^{S} \mathbf{\Gamma}_{BS A \dot{C}}
\end{equation}
where
\begin{eqnarray}
&& \partial_{A \dot{C}} := \sqrt{2} [\partial_{\dot{C}}, \eth_{\dot{C}}] , \ \partial_{\dot{A}} := \frac{\partial}{\partial p^{\dot{A}}} , \ \eth_{\dot{A}} := \phi^{2} \left( \frac{\partial}{\partial q^{\dot{A}}} - Q_{\dot{A}}^{\ \ \dot{B}} \partial_{\dot{B}} \right)
\\ \nonumber
&& Q^{\dot{A} \dot{B}} := -2 \, J^{(\dot{A}} \partial^{\dot{B})} W - \phi \, \partial^{\dot{A}} \partial^{\dot{B}} W
\end{eqnarray}
$\mathbf{\Gamma}_{BS A \dot{C}}$ are spinorial connection coefficients and they read
\begin{eqnarray}
\label{rozpisane_wspolczynniki_koneksji_spinorowej}
&&\mathbf{\Gamma}_{111 \dot{D}} =0  \ \ \ \ \ \ \ \ \ \ \ \ \ \ \ \ \  \ \ \ \ \ \ \ \ \ \ \ \ \ 
\mathbf{\Gamma}_{112 \dot{D}} = -\sqrt{2} \, \phi^{-1} \, J_{\dot{D}}
\\ \nonumber
&&\mathbf{\Gamma}_{121 \dot{D}} = \frac{3}{\sqrt{2}} \phi^{-1} \, J_{\dot{D}} \ \ \ \ \ \ \ \ \ \ \ \ \ \ \ \ \ \ 
  \mathbf{\Gamma}_{122 \dot{D}} = - \frac{1}{\sqrt{2}} \, \phi \, \partial^{\dot{A}} (\phi \, Q_{\dot{A}\dot{D}})
\\ \nonumber
&& \mathbf{\Gamma}_{221 \dot{D}} = \sqrt{2} \phi \, Q_{\dot{D}\dot{A}} J^{\dot{A}}
\ \ \ \ \ \ \ \ \ \ \ \ \ \ \ \,
\mathbf{\Gamma}_{222 \dot{D}} = -\sqrt{2} \phi^{2} \, \eth^{\dot{A}} Q_{\dot{A}\dot{D}}
\\ \nonumber
&&\mathbf{\Gamma}_{\dot{A} \dot{B} 1 \dot{D}} = -\sqrt{2} \phi^{-1} \, J_{(\dot{A}} \in_{\dot{B})\dot{D}}
\ \ \ \ \ \ 
\mathbf{\Gamma}_{\dot{A} \dot{B} 2 \dot{D}} =  \sqrt{2} \, \phi \Big(  
\phi \, \partial_{(\dot{A}} Q_{\dot{B})\dot{D}} +  \in_{\dot{D}(\dot{B}} Q_{\dot{A})\dot{C}} J^{\dot{C}} \Big) 
\end{eqnarray}

In the $\mathcal{HH}$-spaces of the type $[\textrm{N}] \otimes [\textrm{any}]$ the basis of 1-index undotted spinors is chosen in such a manner that 
\begin{equation}
\label{baza_spinorow_niekropkowanych_w_HH}
a_{A}=(0,r), \ r \ne 0
\end{equation} 
From (\ref{rownania_strun_SD_rozpisane}) one finds 
\begin{equation}
\label{zwiazek_spinora_J_z_ekspansja}
M_{\dot{A}} = - \sqrt{2} \phi^{-1} r J_{\dot{A}}
\end{equation}
so the expansion of congruence of SD null strings in expanding $\mathcal{HH}$-spaces is proportional to the spinor field $J_{\dot{A}}$. 

Coordinates $p^{\dot{A}}$ are coordinates on the null strings, while $q^{\dot{A}}$ play the role of parameters labelling the null strings. From (\ref{zwiazaki_miedzy_wspolrzednymi}) we find that $w=J_{\dot{A}} q^{\dot{A}}$ and $t=K^{\dot{A}} q_{\dot{A}}$ what shows that there is a fundamental difference between PRF-coordinates $w$ and $t$. Coordinate $w$ is related to the spinor $J_{\dot{A}}$ and, consequently, to the expansion of the congruence of SD null strings. 

The expanding (nonexpanding) $\mathcal{HH}$-spaces of the SD type [N] will be denoted by $[\textrm{N}]^{e} \otimes [\textrm{any}]$ ($[\textrm{N}]^{n} \otimes [\textrm{any}]$).

\subsection{Homothetic symmetries in $\mathcal{HH}$-spaces of the type $[\textrm{N}]^{e} \otimes [\textrm{any}]$}

Consider the vector field $K$ which satisfies the following set of equations
\begin{equation}
\label{uklad_rownan_Killinga}
\nabla_{(a} K_{b)} = \chi g_{ab}
\end{equation}
If $\chi = 0$ then the vector field $K^{a}$ is called \textsl{the Killing vector field (Killing vector)}. If $\chi = \chi_{0} = \textsl{const}$ then vector field is called \textsl{homothetic vector field (homothetic vector)}. It reduces to Killing vector if $\chi_{0}=0$. If $\chi_{0} \ne 0$ then $K^{a}$ is called the \textsl{proper homothetic vector field (proper homothetic vector}). Finally, if $\nabla_{a} \chi \ne 0$ then $K^{a}$ is \textsl{proper conformal vector} (which is called \textsl{special} if $\nabla_{b} \nabla_{a} \chi =0$ and \textsl{non-special} if $\nabla_{b} \nabla_{a} \chi \ne 0$). Proper conformal vectors in Einstein spaces are rare and they are admitted by the spaces of the type $[\textrm{N},-] \otimes [\textrm{N},-]$ only. Proper conformal vectors are not considered in this paper since they have been analyzed elsewhere (see \cite{Chudecki_Killingi_nieekspandujace} for the type $[\textrm{N}] \otimes [\textrm{N}]$ and \cite{Chudecki_Dobrski} for the type $[\textrm{N}] \otimes [-]$).

Homothetic symmetries in expanding $\mathcal{HH}$-spaces have been widely analyzed in \cite{Finley_Sonnleitner, Chudecki_Killingi}. It was proved that any homothetic vector in $\mathcal{HH}$-spaces of the type $[\textrm{N}]^{e} \otimes [\textrm{any}]$ can be brought to the form 
\begin{eqnarray}
&& K=a \, \frac{\partial}{\partial w} + b \, \frac{\partial}{\partial t} +  (b_{t}-2\chi_{0}) \phi   \, \frac{\partial}{\partial \phi} 
\\ \nonumber
&& \ \ \ \ \ \ \ \ \ \ \ \ \ \ \ \ 
+ \Big( (2b_{t}-a_{w} -2 \chi_{0} ) \eta + b_{w} \phi - \tau \epsilon \Big) \frac{\partial}{\partial \eta}
\end{eqnarray}
Set of ten Eqs. (\ref{uklad_rownan_Killinga}) have been reduced to the single \textsl{master equation}
\begin{eqnarray}
\label{master_equation}
K(W) &=& - (4 \chi_{0} +2a_{w} - 3b_{t} ) W + \alpha \phi^{3}
\\ \nonumber
&& +\frac{1}{2 \tau} \Big( -b_{ww} \phi^{2} - b_{tt} \eta^{2} + (a_{ww} - 2 b_{tw}) \eta \phi \Big) 
+ \frac{1}{2} ( \epsilon_{w} \phi + \epsilon_{t} \eta) + \beta
\end{eqnarray}
where $a=a(w)$ and $b, \alpha, \epsilon$ and $\beta$ are functions of $(w,t)$. Functions $a$, $b$ and structural function $\gamma$ are related by one constraint equation
\begin{equation}
\label{warunek_calkowalnosci}
2a \gamma_{w} + 2 b \gamma_{t} + 4 \gamma a_{w} - a_{www}=0
\end{equation}
Eq. (\ref{warunek_calkowalnosci}) is, in fact, the integrability condition of Eq. (\ref{master_equation}).

Two important remarks should be done here. If the space admits two linearly independent homothetic vectors then without any loss of generality we can assume that, at least, one of them is a Killing vector (the proof of this fact is straightforward, see, e.g. \cite{Chudecki_Przanowski}). 

The second remark concerns the null homothetic vectors. In \cite{Chudecki_null_Killing} spaces equipped with such vectors have been considered in details. It was proved that proper null homothetic vectors are admitted only by the spaces of the type $[\textrm{N}]^{e} \otimes [\textrm{III}]^{n}$, $[\textrm{III}]^{n} \otimes [-]$ and $[\textrm{N}]^{e} \otimes [-]$. Consequently, no spaces of the type $[\textrm{N}] \otimes [\textrm{N}]$ admit proper null homothetic vector. However, spaces of the type $[\textrm{N}] \otimes [\textrm{N}]$ can be equipped with the null Killing vector. These are all spaces of the type $[\textrm{N}]^{n} \otimes [\textrm{N}]^{n}$ (see section \ref{Sekcja_pp_fale}) and special subclass of the spaces of the type $[\textrm{N}]^{n} \otimes [\textrm{N}]^{e}$ (see section \ref{Sekcja_klasa_I_b}).


\setcounter{equation}{0}
\section{Expanding $\mathcal{HH}$-spaces of the type $[\textrm{N}] \otimes [\textrm{N}]$}
\label{section_przestrzenie_NxN}

\subsection{Congruences of ASD null strings and their intersection with congruences of SD null strings}
\label{subsekcja_intersection}

In this section we consider spaces of the type $[\textrm{N}] \otimes [\textrm{N}]$, i.e., we assume, that ASD part of the Weyl tensor is of the type [N]. According to the complex Goldberg - Sachs theorem it follows, that such spaces admit congruence of ASD null strings, determined by the dotted Penrose spinor $a_{\dot{A}}$ with the expansion given by the spinor $M_{C}$
\begin{equation}
\label{rownania_strun_ASD}
a^{\dot{B}} \nabla_{C \dot{A}} a_{\dot{B}} = a_{\dot{A}} M_{C} 
\end{equation}
Eqs. (\ref{rownania_strun_ASD}) can be rewritten in the form
\begin{equation}
\label{rownania_strun_ASD_rozpisane}
a_{\dot{B}} M_{A} = a^{\dot{M}} \partial_{A \dot{B}} a_{\dot{M}} + a^{\dot{M}} a^{\dot{S}} \mathbf{\Gamma}_{\dot{M}\dot{S} A \dot{B}}
\end{equation}
For further purposes it is desired to write SD and ASD null strings equations (\ref{rownania_strun_SD}) and (\ref{rownania_strun_ASD}) in the form
\begin{subequations}
\begin{eqnarray}
&& \nabla_{A\dot{M}} a_{B} = Z_{A\dot{M}} a_{B} + \in_{AB} M_{\dot{M}}
\\ 
&& \nabla_{A\dot{M}} a_{\dot{B}} = \dot{Z}_{A\dot{M}} a_{\dot{B}} + \in_{\dot{M}\dot{B}} M_{A}
\end{eqnarray}
\end{subequations}
where $Z_{A\dot{M}}$ and $\dot{Z}_{A\dot{M}}$ are \textsl{Sommers vectors} \cite{Plebanski_Rozga}.

Both congruences of the null strings mutually intersect and these intersections constitute the congruence of the (complex) null geodesic lines. The null vector tangent to both null strings has the form $K_{A \dot{B}}=ka_{A}a_{\dot{B}}$. We decompose the covariant derivative of the vector $K_{A\dot{B}}$ into its irreducible parts
\begin{equation}
\nabla_{A}^{\ \; \dot{B}} K_{M}^{\ \; \dot{N}} = \Omega_{AM}^{\ \ \ \; \dot{B}\dot{N}} + \frac{1}{2} \in_{AM} l^{\dot{B}\dot{N}} + \frac{1}{2} \in^{\dot{B}\dot{N}} l_{AM} + \frac{1}{4} \in_{AM} \in^{\dot{B}\dot{N}} \xi
\end{equation}
where $\Omega_{AM}^{\ \ \ \; \dot{B}\dot{N}}=\Omega_{(AM)}^{\ \ \ \ \ \dot{B}\dot{N}}=\Omega_{AM}^{\ \ \ \; (\dot{B}\dot{N})}$, $l_{AM}=l_{(AM)}$ and $l^{\dot{B}\dot{N}} = l^{(\dot{B}\dot{N})}$. For the null vector $K_{A\dot{B}} = k a_{A} a_{\dot{B}}$ we easily find
\begin{subequations}
\label{independent_part}
\begin{eqnarray}
\label{independent_part_1}
 \Omega_{AM}^{\ \ \ \; \dot{B}\dot{N}} &=& k a_{(M} a^{(\dot{N}} \mathcal{K}_{A)}^{\ \; \dot{B})}
\\
 l^{\dot{B}\dot{N}} &=& -k a^{A} a^{(\dot{N}} \mathcal{K}_{A}^{\ \; \dot{B})} + 2k a^{(\dot{N}} M^{\dot{B})}
\\
l_{AM} &=& -k a^{\dot{A}} a_{(M} \mathcal{K}_{A)\dot{A}} + 2k a_{(M} M_{A)}
\\
\label{independent_part_4}
 \xi &=& k a^{A} a^{\dot{A}} \mathcal{K}_{A \dot{A}} + 2k (a_{A}M^{A} + a_{\dot{A}}M^{\dot{A}})
\end{eqnarray}
\end{subequations}
where we used abbreviation $\mathcal{K}_{A \dot{B}} := \nabla_{A \dot{B}} \ln k +  Z_{A \dot{B}} +  \dot{Z}_{A \dot{B}}$. Now we define the following quantities
\begin{subequations}
\begin{eqnarray}
\theta &:=& \frac{1}{2} \nabla^{a}K_{a} = - \frac{1}{4} \xi 
\\
\varrho &:=& \frac{1}{2} \nabla_{[a}K_{b]} \, \nabla^{a}K^{b} = \frac{1}{16} (l_{AB}l^{AB} + l_{\dot{A}\dot{B}} l^{\dot{A}\dot{B}} )
\\
s &:=& \frac{1}{2} \nabla_{(a}K_{b)} \, \nabla^{a}K^{b} - \theta^{2} = \frac{1}{8} \Omega_{AB\dot{C}\dot{D}} \Omega^{AB\dot{C}\dot{D}} - \frac{1}{32} \xi^{2}
\end{eqnarray}
\end{subequations}
Scalars $\theta$, $\varrho$ and $s$ describe the invariant properties of the matrix $\nabla_{a}K_{b}$. In real Lorentzian spaces with the null geodesic congruence being in the affine parametrization these scalars have transparent geometrical interpretation. Namely, $\theta$ is the \textsl{expansion of the congruence}, $\varrho=\omega^{2}$ where $\omega$ is the \textsl{twist of the congruence} and $s=\sigma \bar{\sigma}$ where $\sigma$ is \textsl{shear of the congruence}. In complex (real neutral) spaces the interpretation of these parameters is not clear yet, but, analogously as in Lorentzian spaces we call them expansion, twist and shear, respectively, of the congruence of complex (real neutral) geodesics. 

Using (\ref{independent_part}) we find
\begin{subequations}
\label{skalary_ogolniee}
\begin{eqnarray}
\label{ekspansja_ogolna}
\theta &=& -\frac{k}{4} (a_{A}a_{\dot{B}} \mathcal{K}^{A\dot{B}} + 2 a_{A}M^{A} + 2 a_{\dot{A}}M^{\dot{A}})
\\
\varrho &=& -\frac{k^{2}}{16} \Big( (a_{A}a_{\dot{B}} \mathcal{K}^{A\dot{B}})^{2} +2 (a_{A}a_{\dot{B}} \mathcal{K}^{A\dot{B}}) ( a_{A}M^{A} +  a_{\dot{A}}M^{\dot{A}} ) 
\\ \nonumber
&& \ \ \  + 2(a_{A}M^{A})^{2} + 2 (a_{\dot{A}}M^{\dot{A}})^{2} \Big)
\\ 
\label{shear_ogolna}
s &=& -\frac{k^{2}}{8} (a_{A}a_{\dot{B}} \mathcal{K}^{A\dot{B}} +  a_{A}M^{A} +  a_{\dot{A}}M^{\dot{A}}) (a_{A}M^{A} +  a_{\dot{A}}M^{\dot{A}})
\end{eqnarray}
\end{subequations}
Null geodesic congruence with optical properties described by Eqs.(\ref{skalary_ogolniee}) is not presented yet in the affine parametrization. Straightforward computations show that
\begin{equation}
K_{M \dot{N}} \nabla^{M \dot{N}} K_{A\dot{B}} = k K_{A\dot{B}} \Big( a_{M}a_{\dot{M}} \mathcal{K}^{M\dot{M}} + a_{M}M^{M} +  a_{\dot{M}}M^{\dot{M}} \Big)
\end{equation}
so the congruence of the null geodesics is written in the affine parametrization if
\begin{equation}
\label{rownanie_na_parametryzacje_afiniczna}
a_{A}a_{\dot{B}} (\nabla^{A \dot{B}} \ln k +  Z^{A \dot{B}} +  \dot{Z}^{A \dot{B}}) + a_{A}M^{A} +  a_{\dot{A}}M^{\dot{A}} = 0
\end{equation}
Now we assume that the parameter $k$ has been chosen in such a manner, that the equation (\ref{rownanie_na_parametryzacje_afiniczna}) is fulfilled (the congruence of null geodesics is in the affine parametrization). Then Eqs. (\ref{skalary_ogolniee}) simplify considerably and they read
\begin{subequations}
\label{afiniczne_skalary_optyczne}
\begin{eqnarray}
\label{ekspansja_affiniczna}
\theta &=& -\frac{k}{4} ( a_{A}M^{A} +  a_{\dot{A}}M^{\dot{A}})
\\
\label{twist_affiniczna}
\varrho &=& -\frac{k^{2}}{16} \big( a_{A}M^{A} - a_{\dot{A}}M^{\dot{A}} \big)^{2}
\\ 
\label{shear_affiniczna}
s &=& 0
\end{eqnarray}
\end{subequations}
Eqs. (\ref{afiniczne_skalary_optyczne}) determine the relation between properties of the congruences of the null strings ($M_{A}$, $M_{\dot{A}}$) and properties of the congruence of the null geodesics ($\theta$, $\varrho$). Note that the expansion $\theta$ and the twist $\varrho$ do not depend on the Sommers vectors of the congruences of the null strings but only on the expansions $M_{A}$ and $M_{\dot{A}}$.

\subsection{Classification of the spaces of the type $[\textrm{N}] \otimes [\textrm{N}]$}

Using (\ref{baza_spinorow_niekropkowanych_w_HH}), (\ref{zwiazek_spinora_J_z_ekspansja}) and (\ref{afiniczne_skalary_optyczne}) one arrives at the relations which remain valid in all expanding $\mathcal{HH}$-spaces
\begin{subequations}
\label{skalary_optyczne_ogolnie}
\begin{eqnarray}
 \theta &\sim & a_{A}M^{A} + a_{\dot{A}} M^{\dot{A}} = rM_{1} -   \sqrt{2} \phi^{-1} r a_{\dot{A}}J^{\dot{A}}
\\
\varrho &\sim & a_{A}M^{A} - a_{\dot{A}} M^{\dot{A}} = rM_{1} + \sqrt{2} \phi^{-1} r a_{\dot{A}}J^{\dot{A}} 
\end{eqnarray}
\end{subequations}
We propose to gather the properties of the congruence of null geodesics in the affine parametrization by using the following symbol: $[et]$. In this symbol $e$ means expansion, $t$ means twist, $e,t = \{ +,- \}$ and
\begin{eqnarray}
\nonumber
[++]: \, \theta \ne 0, \varrho \ne 0
\\ \nonumber
[+-]: \, \theta \ne 0, \varrho = 0
\\ \nonumber
[-+]: \, \theta = 0, \varrho \ne 0
\\ \nonumber
[--]: \, \theta = 0, \varrho = 0
\end{eqnarray}

At this point we arrive at the crucial idea of this paper. Complex spaces equipped with the congruences of SD and ASD  null strings can be classified according to properties of these congruences (expanding, nonexpanding) and the properties of their intersection ($[++], [+-], [-+], [--]$). Vacuum type $[\textrm{N}] \otimes [\textrm{N}]$ spaces can be divided into six subtypes, see Table \ref{Tabela_typy_N}.
\renewcommand{\arraystretch}{1,2}
\begin{table}[ht]
\begin{center}
\begin{tabular}{|c|c|}   \hline
Type & considered in  \\  \hline
$\{ [\textrm{N}]^{n} \otimes [\textrm{N}]^{n},[--] \}$ & section \ref{Sekcja_pp_fale}        \\  \hline
$\{ [\textrm{N}]^{e} \otimes [\textrm{N}]^{n},[--] \}$ & section \ref{Sekcja_klasa_I_b}              \\  \hline
$\{ [\textrm{N}]^{e} \otimes [\textrm{N}]^{n},[++] \}$ & section \ref{Sekcja_klasa_III_szczegolna} \\  \hline
$\{ [\textrm{N}]^{e} \otimes [\textrm{N}]^{e},[--] \}$ & section \ref{Sekcja_klasa_I}              \\  \hline
$\{ [\textrm{N}]^{e} \otimes [\textrm{N}]^{e},[+-] \}$ & section \ref{Sekcja_klasa_II}             \\  \hline
$\{ [\textrm{N}]^{e} \otimes [\textrm{N}]^{e},[++] \}$ & see \cite{Chudecki_Przanowski}     \\  \hline
\end{tabular}
\caption{Possible types of $[\textrm{N}] \otimes [\textrm{N}]$ spaces.}
\label{Tabela_typy_N}
\end{center}
\end{table}

\textbf{Remark}. Although we consider only vacuum spaces, such classification can be considered also in non-vacuum spaces. In non-vacuum spaces one more type, namely $\{ [\textrm{N}]^{e} \otimes [\textrm{N}]^{e},[-+] \}$, appears. From the Raychaudhuri equation it follows that the type $\{ [\textrm{N}]^{e} \otimes [\textrm{N}]^{e},[-+] \}$ does not exist in the vacuum case.

\subsection{Key function for the type $[\textrm{N}] \otimes [\textrm{N}]$}

The fact that ASD curvature is of the type [N] imposes strong restrictions on the key function $W$. ASD conformal curvature (\ref{ASD_conformal_curvature_ogolnie}) can be written in the form
\begin{eqnarray}
\label{ASD_curvatura_rozpisana}
C_{\dot{A}\dot{B}\dot{C}\dot{D}} &=& \phi^{3} ( J_{(\dot{A}} J_{\dot{B}} J_{\dot{C}} J_{\dot{D})} W_{\phi \phi \phi \phi}-4 J_{(\dot{A}} J_{\dot{B}} J_{\dot{C}} K_{\dot{D})} W_{\phi \phi \phi \eta} 
\\ \nonumber
&&+ 6 J_{(\dot{A}} J_{\dot{B}} K_{\dot{C}} K_{\dot{D})} W_{\phi \phi \eta \eta} - 4 J_{(\dot{A}} K_{\dot{B}} K_{\dot{C}} K_{\dot{D})} W_{\phi \eta \eta \eta} + K_{(\dot{A}} K_{\dot{B}} K_{\dot{C}} K_{\dot{D})} W_{\eta \eta \eta \eta} )
\end{eqnarray}
Contraction of (\ref{ASD_curvatura_rozpisana}) with $\xi^{\dot{A}} \xi^{\dot{B}} \xi^{\dot{C}} \xi^{\dot{D}}$ (where $\xi^{\dot{A}}$ is an arbitrary nonzero spinor) gives
\begin{eqnarray}
\mathcal{C} (x,y) &=& C_{\dot{A}\dot{B}\dot{C}\dot{D}} \xi^{\dot{A}} \xi^{\dot{B}} \xi^{\dot{C}} \xi^{\dot{D}}
\\ \nonumber
 &=& \phi^{3} ( x^{4} W_{\phi \phi \phi \phi}
+4x^{3}y W_{\phi \phi \phi \eta} + 6 x^{2} y^{2} W_{\phi \phi \eta \eta} +4xy^{3} W_{\phi \eta \eta \eta} + y^{4} W_{\eta \eta \eta \eta})
\end{eqnarray}
where $x := J_{\dot{A}} \xi^{\dot{A}}$, $y := -K_{\dot{A}} \xi^{\dot{A}}$. Hence $\mathcal{C}$ is fourth order polynomial in $x$ and $y$. ASD curvature is of the type [N] if the polynomial $\mathcal{C}$ has one quadruple root. Hence, it can be written in the form $\mathcal{C} (x,y) =(f_{1} y+ f_{2} x)^{4}$.

The first possibility is $W_{\eta \eta \eta \eta}=0$. This implies $W_{\phi \eta \eta \eta}=W_{\phi \phi \eta \eta}=W_{\phi \phi \phi \eta}=0$. ASD curvature has the form 
\begin{equation}
C_{\dot{A}\dot{B}\dot{C}\dot{D}} = \phi^{3} W_{\phi \phi \phi \phi} J_{\dot{A}} J_{\dot{B}} J_{\dot{C}} J_{\dot{D}}
\end{equation}
Consequently, $W_{\phi \phi \phi \phi} \ne 0$ and 4-fold Penrose spinor is $a_{\dot{A}}=J_{\dot{A}}$ (see subsection \ref{subsekcja_Klasa_I} for details).

The second possibility is given by $W_{\eta \eta \eta \eta} \ne 0$. We obtain
\begin{eqnarray}
\label{rownania_na_typ_N_ASD}
&& W_{\eta \eta \eta \phi} = h W_{\eta \eta \eta \eta}
\\ \nonumber
&& W_{\eta \eta \phi \phi} = h W_{\eta \eta \eta \phi}
\\ \nonumber
&& W_{\eta \phi \phi \phi} = h W_{\eta \eta \phi \phi}
\\ \nonumber
&& W_{\phi \phi \phi \phi} = h W_{\eta \phi \phi \phi}
\end{eqnarray}
where $h=h(\phi, \eta, w,t)$. ASD curvature has the form
\begin{equation}
\label{ASD_krzywizna_przy_Weta4ne0}
C_{\dot{A}\dot{B}\dot{C}\dot{D}} = \phi^{3} W_{\eta \eta \eta \eta} a_{\dot{A}} a_{\dot{B}} a_{\dot{C}} a_{\dot{D}} \ , \ \ \ a_{\dot{A}} := hJ_{\dot{A}} - K_{\dot{A}}
\end{equation}
Using (\ref{transformacja_funkcji_kluczowej}) and (\ref{rownania_na_typ_N_ASD}) one finds the transformation law for  $h$
\begin{equation}
\label{transformacja_na_h}
w'_{w} \, h' = \lambda^{-\frac{1}{2}} h - t'_{w}
\end{equation}
The integrability conditions of the system (\ref{rownania_na_typ_N_ASD}) lead to only one equation
\begin{equation}
\label{rownanie_na_h}
h_{\phi} = h h_{\eta}
\end{equation}
Further steps depend on $h_{\eta}$. If $h_{\eta} = 0$ then $h_{\phi}=0$ and finally $h=h(w,t)$. In this case function $h$ can be gauged to zero, $h=0$ (compare (\ref{transformacja_na_h})). It implies $W_{\phi \eta \eta \eta}=W_{\phi \phi \eta \eta}=W_{\phi \phi \phi \eta}=W_{\phi \phi \phi \phi}=0$. ASD curvature takes the form
\begin{equation}
C_{\dot{A}\dot{B}\dot{C}\dot{D}} = \phi^{3} W_{\eta \eta \eta \eta} K_{\dot{A}} K_{\dot{B}} K_{\dot{C}} K_{\dot{D}}
\end{equation}
with 4-fold Penrose spinor $a_{\dot{A}}=K_{\dot{A}}$. This case is considered in details in the subsection \ref{subsekcja_Klasa_II}.

The most general case is characterized by $W_{\eta \eta \eta \eta} \ne 0$ and $h_{\eta} \ne 0$. Solution of the Eq. (\ref{rownanie_na_h}) is given by the formula
\begin{equation}
\label{rozwiazanie_na_h}
\eta + \phi h = f(h,w,t)
\end{equation}
where $f=f(h,w,t)$ is an arbitrary function. This case is considered in the subsection \ref{subsekcja_Klasa_III}.

\subsection{Case $W_{\eta \eta \eta \eta}=0$}
\label{subsekcja_Klasa_I}

\subsubsection{Key function and properties of the congruences}

In this case we have $W_{\eta \eta \eta \eta}=W_{\phi \eta \eta \eta}=W_{\phi \phi \eta \eta}=W_{\phi \phi \phi \eta}=0$ i $W_{\phi \phi \phi \phi} \ne 0$. 4-fold Penrose spinor is $J_{\dot{A}}$. From (\ref{rownania_strun_ASD_rozpisane}) one gets $W_{\eta \eta \eta}=0$ and
\begin{equation}
M_{1} = 0 , \ M_{2} = -\sqrt{2} \tau^{2} \phi^{3} W_{\eta \eta \phi}
\end{equation}
Obviously $\theta = \varrho = 0$ (compare (\ref{skalary_optyczne_ogolnie})). The general form of the key function reads
\begin{equation}
\label{funkcja_kluczowa_klasa_I}
W(\phi, \eta, w, t) = F(\phi , w, t) + A \phi \eta^{2} + B \eta \phi^{2} + C \eta^{2} + n \eta \phi + m \eta , \ \ F_{\phi\phi\phi\phi} \ne 0
\end{equation}
where $A,B,C,n,m$ are arbitrary functions of $(w,t)$ only. Expansion of the congruence of ASD null strings depends on $W_{\eta \eta \phi} = 2A$, so the possible types are 
\begin{eqnarray}
\nonumber
\{ [\textrm{N}]^{e} \otimes [\textrm{N}]^{e},[--] \} \textrm{  if  } A \ne 0
\\ \nonumber
\{ [\textrm{N}]^{e} \otimes [\textrm{N}]^{n},[--] \} \textrm{  if  } A = 0
\end{eqnarray}

\subsubsection{Field equations}

Inserting (\ref{funkcja_kluczowa_klasa_I}) into the hyperheavenly equation (\ref{rownanie_HH}) one arrives at the system of equations
\begin{subequations}
\label{rownania_pola_klasa_I}
\begin{eqnarray}
\label{rownania_pola_klasa_I_row_1}
&& 4\tau AC-A_{t}=0
\\ 
\label{rownania_pola_klasa_I_row_2}
&& 2 \tau CB + A_{w} - B_{t}=0
\\
\label{rownania_pola_klasa_I_row_3}
&& 4 \tau Am +2 C_{w} - n_{t} = 0
\\
\label{rownania_pola_klasa_I_row_4}
&& (2A\phi + 2C) (\phi F_{\phi \phi} - 2 F_{\phi}) + (2B \phi +n)(n \phi +2m) 
\\ \nonumber
&& \ \ \ \ \ \ \ \ + \frac{1}{\tau} (B_{w} \phi^{2} + n_{w} \phi + m_{w} - F_{\phi t}) - \frac{\gamma}{\tau^{2}} \phi = 0
\end{eqnarray}
\end{subequations}
Field equations (\ref{rownania_pola_klasa_I}) can be completely solved (see sections \ref{Sekcja_klasa_I} i \ref{Sekcja_klasa_I_b}).

\subsection{Case $W_{\eta \eta \eta \eta} \ne 0$, $h=0$}
\label{subsekcja_Klasa_II}

\subsubsection{Key function and properties of the congruences}

Here we have $W_{\phi \phi \phi \phi} = W_{\phi \phi \phi \eta} = W_{\phi \phi \eta \eta} = W_{\phi \eta \eta \eta} = 0$ and $W_{\eta \eta \eta \eta} \ne 0$, the 4-fold Penrose spinor is $K_{\dot{A}}$. From (\ref{rownania_strun_ASD_rozpisane}) one finds $W_{\phi \phi} - \phi W_{\phi \phi \phi}=0$ and
\begin{eqnarray}
&& M_{1} = \sqrt{2} \tau \phi^{-1}
\\ \nonumber
&& M_{2} = \sqrt{2} \tau^{2} \phi ( \phi^{2} W_{\phi \phi \eta} + W_{\eta} - \phi W_{\phi \eta} )
\end{eqnarray}
Since $M_{1} \ne 0$ the congruence of ASD null strings is always expanding. Moreover from (\ref{skalary_optyczne_ogolnie}) it follows that $\theta \ne 0$ and $\varrho = 0$. This implies the space of the type
\begin{equation}
\nonumber
\{ [\textrm{N}]^{e} \otimes [\textrm{N}]^{e},[+-] \} 
\end{equation}
Key function has the form
\begin{equation}
\label{funkcja_kluczowa_dla_NexNe_plusminus}
W(\phi, \eta, w, t) = F(\eta , w, t) + A \phi^{3} + B \phi \eta^{2}  + C \eta \phi + m  \phi  , \ \ F_{\eta\eta\eta\eta} \ne 0
\end{equation}
where $A,B,C,m$ are arbitrary functions of $(w,t)$.

\subsubsection{Field equations}

From the hyperheavenly equation (\ref{rownanie_HH}) we find
\begin{subequations}
\label{rownania_pola_typNexNe_plusminus}
\begin{eqnarray}
&& A=A(w) 
\\
&&  B=B(t)
\\
&& \frac{\gamma}{\tau^{2}} = C^{2} + \frac{1}{\tau} C_{w} - 4Bm
\\ 
&& 2(B \eta^{2} + C \eta +m) F_{\eta \eta} - 2(2B \eta +C) F_{\eta} - \frac{1}{\tau} F_{\eta w} + \frac{1}{\tau} (B_{t} \eta^{2} + C_{t} \eta +m_{t})=0 \ \ \ \ \ \ \ \ \ \ \
\end{eqnarray}
\end{subequations}
Such spaces are considered in section \ref{Sekcja_klasa_II}.

\subsection{Case $W_{\eta \eta \eta \eta} \ne 0$, $h \ne 0$}
\label{subsekcja_Klasa_III}

\subsubsection{Key function and properties of the congruences}

In this case the key function satisfies the system of equations (\ref{rownania_na_typ_N_ASD}) together with the condition $W_{\eta \eta \eta \eta} \ne 0$. 4-fold Penrose spinor is $a_{\dot{A}} = hJ_{\dot{A}} - K_{\dot{A}}$. The equations of ASD null strings $a^{\dot{A}} a^{\dot{B}} \nabla_{A\dot{A}} a_{\dot{B}}$ read
\begin{subequations}
\label{rownania_struny_ASD}
\begin{eqnarray}
\label{pierwsze_rownanie_struny_ASD}
&& h_{\phi} - hh_{\eta}=0
\\ 
\label{drugie_rownanie_struny_ASD}
&& \frac{1}{\tau} (h_{w} - hh_{t}) + \phi (W_{\phi \phi \phi} - 3h W_{\phi \phi \eta} + 3h^{2} W_{\phi \eta \eta} - h^{3}W_{\eta \eta \eta} )
\\ \nonumber
&& \ \ \ \ \ \ \ \ + (\phi h_{\eta} -1) (W_{\phi \phi} - 2hW_{\phi \eta} + h^{2} W_{\eta \eta} ) + 2 h_{\eta} (h W_{\eta} - W_{\phi}) = 0
\end{eqnarray}
\end{subequations}
[\textbf{Remark}. Note, that Eq. (\ref{pierwsze_rownanie_struny_ASD}) is exactly the integrability condition (\ref{rownanie_na_h}) of the Eqs. (\ref{rownania_na_typ_N_ASD})].

Expansion of the congruence of ASD null strings takes the form
\begin{subequations}
\label{ekspansje_struny_ASD}
\begin{eqnarray}
\label{ekspansje_struny_ASD_M1}
M_{1} &=& -\tau \sqrt{2} (h_{\eta} + \phi^{-1})
\\ 
\label{ekspansje_struny_ASD_M2}
M_{2} &=& -\tau \sqrt{2} \phi^{2} h_{t} +\tau^{2} \sqrt{2} \phi (1- \phi h_{\eta}) (\phi W_{\phi \eta} - W_{\eta} - h \phi W_{\eta \eta}) 
\\ \nonumber
&&  -\tau^{2} \sqrt{2} \phi^{3} (h^{2} W_{\eta \eta \eta}  -2h W_{\phi \eta \eta} + W_{\phi \phi \eta} )
\end{eqnarray}
\end{subequations}
and the parameters of the congruence of null geodesics can be easily determined to be
\begin{eqnarray}
\label{skalary_optyczne}
\theta &\sim &  -\sqrt{2} \tau r (h_{\eta} + 2\phi^{-1})
\\
\varrho &\sim &  -\sqrt{2} \tau r h_{\eta} 
\end{eqnarray}

If $\varrho = 0$ then $h_{\eta}=0 \ \Longrightarrow h_{\phi} = 0$. This implies $h=h(w,t)$ and finally $h=0$ (compare transformation formula (\ref{transformacja_na_h})) what contradicts the basic assumption of this subsection, namely $h \ne 0$. Assume now, that expansion $\theta=0$. From the Raychaudhuri equation it follows, that in Einstein spaces the congruence of null geodesics which is nonexpanding and shearfree is necessarily nontwisting. Consequently, both twist $\varrho$ and expansions $\theta$ are nonzero in this case. The only possible types are 
\begin{eqnarray}
\nonumber
\{ [\textrm{N}]^{e} \otimes [\textrm{N}]^{e},[++] \} 
\\ \nonumber
\{ [\textrm{N}]^{e} \otimes [\textrm{N}]^{n},[++] \} 
\end{eqnarray}

We do not consider the type $\{ [\textrm{N}]^{e} \otimes [\textrm{N}]^{e},[++] \} $ in this paper. Thorough analysis of such spaces equipped with two homothetic vectors has been gathered in \cite{Chudecki_Przanowski}. The type $\{ [\textrm{N}]^{e} \otimes [\textrm{N}]^{e},[++] \} $ with one symmetry and without any symmetries is now intensively analyzed but the results will be presented elsewhere.

In this paper we focus on the type $\{ [\textrm{N}]^{e} \otimes [\textrm{N}]^{n},[++] \} $. From (\ref{pierwsze_rownanie_struny_ASD}) and $M_{1}=0$ the solution for $h$ can be easily found
\begin{equation}
h = \frac{s(w,t)-\eta}{\phi}
\end{equation}
From (\ref{transformacja_na_h}) we find transformation law for $s$
\begin{equation}
w'_{w} \, s' = \lambda^{-1} s + \tau w'_{w} \sigma
\end{equation}
Obviously, $s$ can be gauged away without any loss of generality. Consequently, $h = - \eta / \phi$ and this corresponds to the case $f=0$ in (\ref{rozwiazanie_na_h})). Eqs. (\ref{rownania_na_typ_N_ASD}) can be written in the form
\begin{eqnarray}
&& W_{\eta \eta \eta \phi} = h W_{\eta \eta \eta \eta} \ \Longrightarrow \  
(\phi W_{\phi} + \eta W_{\eta} - 3W)_{\eta \eta \eta} = 0
\\ \nonumber
&& W_{\eta \eta \phi \phi} = h W_{\eta \eta \eta \phi} \ \Longrightarrow \
(\phi W_{\phi} + \eta W_{\eta} - 3W)_{\eta \eta \phi} = 0
\\ \nonumber
&& W_{\eta \phi \phi \phi} = h W_{\eta \eta \phi \phi} \ \Longrightarrow \ 
(\phi W_{\phi} + \eta W_{\eta} - 3W)_{\eta \phi \phi} = 0
\\ \nonumber
&& W_{\phi \phi \phi \phi} = h W_{\eta \phi \phi \phi} \ \Longrightarrow \ 
(\phi W_{\phi} + \eta W_{\eta} - 3W)_{\phi \phi \phi} = 0
\end{eqnarray}
Hence, $\phi W_{\phi} + \eta W_{\eta} - 3W$ is a second order polynomial in $\phi$ and $\eta$ with coefficients depending on $(w,t)$ only. This implies $W(\phi, \eta, w,t) = \phi^{3} \, T(x,w,t) + \mathcal{W}$ where $\mathcal{W}$ is a second order polynomial in $\eta$ and $\phi$ with $(w,t)$-depending coefficients and $x := {\eta} / {\phi}$. The only equations which remain to be solved are (\ref{drugie_rownanie_struny_ASD}) and $M_{2}=0$. From these equations it follows that linear factors in $\mathcal{W}$ vanish so, finally, the key function has the form
\begin{equation}
\label{funkcja_kluczowa_dla_NexNn}
W(\phi, \eta, w,t) = \phi^{3} \, T(x,w,t) + \frac{1}{2} A \eta^{2} + B \eta \phi - \frac{C}{2\tau} \phi^{2}  +D , \ \ x:=\frac{\eta}{\phi}
\end{equation}
where $A$, $B$, $C$ and $D$ are arbitrary functions of $(w,t)$.

\subsubsection{Field equations}

The hyperheavenly equation (\ref{rownanie_HH}) under (\ref{funkcja_kluczowa_dla_NexNn}) splits into the set of equations
\begin{subequations}
\begin{eqnarray}
\label{NexNn_e1}
&& A_{w} - B_{t} = 0
\\
\label{NexNn_e2}
&& \frac{\gamma}{\tau^{2}} = B^{2} +  \frac{AC}{ \tau} + \frac{B_{w}}{\tau} + \frac{C_{t}}{\tau^{2}} 
\\
\label{NexNn_e3}
&& (-Ax^{2} - 2Bx +  \tau^{-1}C ) T_{xx} + (2B+2Ax) T_{x} + \frac{1}{\tau} (T_{xw}-3T_{t}+xT_{xt})=0 \ \ \ \ \ \ \ \
\end{eqnarray}
\end{subequations}
The spaces $\{ [\textrm{N}]^{e} \otimes [\textrm{N}]^{n},[++] \} $ are investigated in section \ref{Sekcja_klasa_III_szczegolna}.


\setcounter{equation}{0}
\section{Type $\{ [\textrm{N}]^{n} \otimes [\textrm{N}]^{n},[--] \}$}
\label{Sekcja_pp_fale}

Type $\{ [\textrm{N}]^{n} \otimes [\textrm{N}]^{n},[--] \}$ is characterized by the existence of SD and ASD congruences of the null strings which are both nonexpanding. They intersect along nonexpanding and nontwisting congruence of complex null geodesics. To obtain this type one should consider nonexpanding $\mathcal{HH}$-space as a generic space and equip this space with additional structure given by the nonexpanding congruence of ASD null strings. Such spaces have not been described in section \ref{Przestrzenie_Nxany_preliminaries}. However, the type $\{ [\textrm{N}]^{n} \otimes [\textrm{N}]^{n},[--] \}$ has been considered in \cite{Chudecki_klasyfikacja_Killingow_HH_nieeks,Chudecki_null_Killing}. For completeness we remind here that the metric of such space can be brought to the form
\begin{equation}
\label{zespolone_pp_fale}
ds^{2} = 2(dydp-dxdq + (N(p,q)+H(y,q)) \, dq^{2} ), \ N_{pp} \ne 0 , \ H_{yy} \ne 0
\end{equation}
Such spaces admit automatically null Killing vector $K_{0} = \partial_{x}$. In \cite{Chudecki_klasyfikacja_Killingow_HH_nieeks} Lorentzian slice of the metric (\ref{zespolone_pp_fale}) has been found. First, introduce complex coordinate transformation, namely
\begin{equation}
q=:u, \ x=:v, \ p=:\zeta , \ y=:\bar{\zeta} , \ N=:f , \ H= :\bar{f}
\end{equation}
The metric (\ref{zespolone_pp_fale}) takes the form 
\begin{equation}
\label{rzeczywiste_pp_fale}
ds^{2} = 2 (d \zeta d \bar{\zeta} - dvdu + (f(\zeta,u) + \bar{f} (\bar{\zeta},u) ) \, du^{2} )
\end{equation}
At this stage the metric (\ref{rzeczywiste_pp_fale}) is still the complex metric and bar does not mean anything. However, if we consider the coordinates $(u,v)$ as the real ones and coordinates $(\zeta, \bar{\zeta})$ as the complex ones (where bar denotes complex conjugation), then the metric (\ref{rzeczywiste_pp_fale}) becomes real metric of Lorentzian signature. It is exactly the famous \textsl{pp-wave metric}, so the metric (\ref{zespolone_pp_fale}) can be called \textsl{complex pp-wave metric}. 

We do not consider additional homothetic or Killing symmetries admitted by the metric (\ref{zespolone_pp_fale}). The real Lorentzian case has been analyzed in details elsewhere (see \cite{Exacty} and references therein). We only mention here, that the space of the type $\{ [\textrm{N}]^{n} \otimes [\textrm{N}]^{n},[--] \}$ is the only non-conformally or non-half-conformally flat space which admits a proper conformal vector (see \cite{Chudecki_Killingi_nieekspandujace} for complex case).


\setcounter{equation}{0}
\section{Type $\{ [\textrm{N}]^{e} \otimes [\textrm{N}]^{e},[--] \}$}
\label{Sekcja_klasa_I}

In this section we analyze the spaces of type $\{ [\textrm{N}]^{e} \otimes [\textrm{N}]^{e},[--] \}$. The metric is generated by the key function in the form (\ref{funkcja_kluczowa_klasa_I}) with $A \ne 0$ and it reads
\begin{eqnarray}
\label{metryka_typu_NexNe_minusminus_ogolna}
ds^{2} &=& 2\phi^{-2} \big\{ \tau^{-1} (d \eta  d w - d \phi  dt) -    \phi (2A \phi +2C) \,   
dt^{2} \ \ \ \ \ \ 
\\ \nonumber
&& \ \ \ \ \ \ \ \ \ \ - ( 2B \phi^{2} - 4C \eta -2m ) \, dw dt
\\ \nonumber
&& \ \ \ \ \ \ \ \ \ \
  +  ( 2F_{\phi} - \phi F_{\phi \phi} + 2A \eta^{2} + 2B \eta \phi + 2n \eta  ) \, dw^{2} \big\} \ \ \ \ \ \ 
\end{eqnarray}

\subsection{Generic case; no symmetries are assumed}
\label{Subsekcja_klasa_I_bez_symetrii}

From the Eq. (\ref{transformacja_funkcji_kluczowej}) we find transformation formulas for the functions $A,B,C,m$ and $n$. With the help of the field equations (\ref{rownania_pola_klasa_I_row_1}-\ref{rownania_pola_klasa_I_row_3}) one finds that without any loss of generality one can put $A=1$, $B=C=m=n=0$ so the key function reads
\begin{equation}
\label{funkcja_kluczowa_klasa_I_uproszczona}
W(\phi, \eta, w, t) = F(\phi , w, t) +  \phi \eta^{2}  , \ \ F_{\phi\phi\phi\phi} \ne 0
\end{equation}
The last remaining field equation (\ref{rownania_pola_klasa_I_row_4}) reduces to the equation 
\begin{equation}
2 \phi (\phi F_{\phi \phi} - 2 F_{\phi}) - \frac{1}{\tau} F_{\phi t} - \frac{\gamma}{\tau^{2}} \phi = 0
\end{equation}
with the solution 
\begin{eqnarray}
&& F_{\phi} = \phi^{2} (g+f)  - \frac{f_{t} \phi}{2 \tau} ,  \ \gamma =: \frac{1}{2} f_{tt}
\\ \nonumber
&& g = g (x,w), \ g_{xxx} \ne 0, \ x:= t - \frac{1}{2 \tau \phi} , \ f=f(w,t) , \ f_{ttt} \ne 0
\end{eqnarray}
Finally, the metric reads
\begin{equation}
\label{metryka_NexNe_min_min}
ds^{2} = 2 \phi^{-2} \left\{ \frac{1}{\tau} (d \eta dw - d \phi dt)  -2 \phi^{2} dt^{2} + \left( 2\eta^{2}  - (g_{x} + f_{t}) \frac{\phi}{2 \tau} \right) dw^{2} \right\}
\end{equation}

The metric (\ref{metryka_NexNe_min_min}) admits a real Lorentzian slice. First, we introduce complex transformation of the variables
\begin{equation}
w=:u, \ t = :\frac{\zeta}{\sqrt{2}}, \ x=:- \frac{\bar{\zeta}}{\sqrt{2}}, \ \eta =: - \frac{v}{2 \tau (\zeta + \bar{\zeta})^{2}}, \ f_{t} =: \sqrt{2} \, H(u, \zeta) , \ g_{x} =: \sqrt{2} \, \bar{H} (u, \bar{\zeta})
\end{equation}
Hence
\begin{eqnarray}
\label{metryka_NexNe_minus_minus_ciecie_Lorent}
ds^{2} &=& 2 d \zeta d \bar{\zeta} - 2 du \left( dv - \frac{2v}{\zeta + \bar{\zeta}} \, d (\zeta + \bar{\zeta}) \right)
\\ \nonumber
&& + 2 \left( \frac{v^{2}}{(\zeta + \bar{\zeta})^{2}} - (\zeta + \bar{\zeta}) (H(u, \zeta) + \bar{H} (u, \bar{\zeta}) ) \right)  du^{2}
\end{eqnarray}
If we now consider the coordinates $(u,v)$ as real ones and the coordinates $(\zeta, \bar{\zeta})$ as complex (where bar denotes the complex conjugation) then the metric (\ref{metryka_NexNe_minus_minus_ciecie_Lorent}) becomes real metric with Lorentzian signature and it is well known as a metric belonging to the \textsl{Kundt class} \cite{Exacty}. Kundt class is characterized by the existence of the nonexpanding and nontwisting congruence of null geodesics, and pp-waves belong to this class as a special subclass. The property which distinguishes these two kind of spaces is the existence of the null Killing vector. Kundt class does not admit such a symmetry while pp-wave class does.

Our considerations prove that there is some deep difference between pp-waves and Kundt class on the level of complexification of these spaces. The generic complex spaces of both these classes are equipped with nonexpanding and nontwisting congruence of null geodesics. The essential difference lies in the properties of the congruences of the null strings. In the case of pp-waves both SD and ASD congruences are nonexpanding, while in the case of the Kundt class both congruences are expanding. Hence
\begin{eqnarray}
\nonumber
 \{ [\textrm{N}]^{n} \otimes [\textrm{N}]^{n},[--] \} & \stackrel{\textrm{real Lorentzian slice}}{\longrightarrow} & \textrm{pp-waves}
\\ \nonumber
 \{ [\textrm{N}]^{e} \otimes [\textrm{N}]^{e},[--] \} & \stackrel{\textrm{real Lorentzian slice}}{\longrightarrow} & \textrm{Kundt class}
\end{eqnarray}

\subsection{One symmetry}
\label{Sekcja_klasa_I_one_symmetry}

Inserting the key function in the form (\ref{funkcja_kluczowa_klasa_I}) into the master equation (\ref{master_equation}) we find the equations which relate the metric functions $A,B,C,m,n$ with functions $a,b,\epsilon,\alpha, \beta$. They read
\begin{subequations}
\label{rownanie_master_dla_[N]ex[N]e_minus_minus}
\begin{eqnarray}
\label{rownanie_master_dla_[N]ex[N]e_minus_minus_1}
&& aA_{w} + bA_{t} + (2b_{t} - 2\chi_{0})A= 0
\\ 
\label{rownanie_master_dla_[N]ex[N]e_minus_minus_2}
&& aC_{w} + bC_{t} +b_{t} C + \frac{1}{2 \tau} b_{tt} = 0
\\ 
\label{rownanie_master_dla_[N]ex[N]e_minus_minus_3}
&& aB_{w} + bB_{t} + (b_{t} + a_{w} - 2 \chi_{0}) B + 2b_{w}A=0
\\ 
\label{rownanie_master_dla_[N]ex[N]e_minus_minus_4}
&& a n_{w} + b n_{t} +a_{w} n + 2b_{w}C - 2 \tau \epsilon A - \frac{1}{2 \tau} (a_{ww} - 2b_{tw}) = 0
\\ 
\label{rownanie_master_dla_[N]ex[N]e_minus_minus_5}
&& am_{w} + bm_{t} + (a_{w} - b_{t} +2 \chi_{0}) m -2 \tau \epsilon C  - \frac{1}{2} \epsilon_{t} = 0
\\ 
\label{rownanie_master_dla_[N]ex[N]e_minus_minus_6}
&& aF_{w} + bF_{t} + (b_{t} - 2\chi_{0}) \phi F_{\phi} + b_{w} \phi (B\phi^{2} + n\phi +m)  - \tau \epsilon (B\phi^{2} + n\phi +m) \ \ \ \ \ \ \ \ \ \ 
\\ \nonumber
&&
 \ \ \ \ \ \ \ \ + (4\chi_{0} + 2a_{w} - 3b_{t}) F - \alpha \phi^{3} + \frac{1}{2 \tau} b_{ww} \phi^{2} - \frac{1}{2} \epsilon_{w} \phi  - \beta = 0
 \end{eqnarray}
 \end{subequations}
We do not use the reduced form of the key function (\ref{funkcja_kluczowa_klasa_I_uproszczona}) because our priority is to bring the homothetic vector to the simplest possible form. To do this we need the gauge freedom which in the case with no symmetries assumed has been used to bring the key function to the form (\ref{funkcja_kluczowa_klasa_I_uproszczona}). This is why we take as a starting point the general form of the key function (\ref{funkcja_kluczowa_klasa_I}). After some work one finds, that homothetic vector can be brought to the form with $a=1$, $b=\epsilon=0$. Namely
\begin{equation}
K_{1} = \partial_{w} - 2 \chi_{0} \left( \phi \partial_{\phi} + \eta \partial_{\eta} \right)
\end{equation}
Using gauge freedom and Eqs. (\ref{rownanie_master_dla_[N]ex[N]e_minus_minus_1}-\ref{rownanie_master_dla_[N]ex[N]e_minus_minus_5}) and (\ref{rownania_pola_klasa_I_row_1}-\ref{rownania_pola_klasa_I_row_3})
one finds the solutions
\begin{equation}
A(w) = e^{2\chi_{0}w}, \ B(w,t) = (2\chi_{0}t + B_{0})e^{2\chi_{0}w}, \ C=m=n=0
\end{equation}
From the integrability condition (\ref{warunek_calkowalnosci}) it follows that $\gamma = \gamma(t)$. Eq. (\ref{rownanie_master_dla_[N]ex[N]e_minus_minus_6}) has the solution 
\begin{equation}
\label{rozwiazanie_naF_NexNe_minusminus_jedna_symetria}
F(\phi, w, t) = e^{-4\chi_{0}w} H(y, t) + \tilde{\alpha} (w,t) \phi^{3} , \ y:= \phi e^{2\chi_{0}w}
\end{equation}
where $\tilde{\alpha}$ is an arbitrary function. The last equation which remains to be solved is the field equation (\ref{rownania_pola_klasa_I_row_4}) which takes the form
\begin{equation}
\label{ostateczne_rownanie_pola_NexNe_minusminus_jedna_symetria}
2(yH_{yy} - 2H_{y})  - \frac{1}{\tau y} H_{ty} + \frac{2\chi_{0}}{\tau} (2 \chi_{0} t +B_{0}) y -\frac{3\tilde{\alpha}_{t}}{\tau} \phi -\frac{\gamma(t)}{\tau^{2}} = 0
\end{equation}
Differentiation of the Eq. (\ref{ostateczne_rownanie_pola_NexNe_minusminus_jedna_symetria}) with respect to $w$ tells us that $\tilde{\alpha} (w,t) = f_{1} (t) e^{2\chi_{0}w} + f_{2} (w)$. Function $f_{2}$ can be gauged away with the help of the gauge function $L(w)$. With $f_{2}=0$ the function $f_{1}$ can be absorbed into $H(y,t)$ (compare (\ref{rozwiazanie_naF_NexNe_minusminus_jedna_symetria})). Hence, $\tilde{\alpha}$ can be put zero without any loss of generality. Solution of the Eq. (\ref{ostateczne_rownanie_pola_NexNe_minusminus_jedna_symetria}) reads
\begin{eqnarray}
\label{rozwiazanie_master_dla_[N]ex[N]e_minus_minus}
&&H_{y} = y^{2} (g + f) + 2\chi_{0} (\chi_{0}t^{2} + B_{0}t) y^{2}  - \frac{f_{t}}{2 \tau} y , \ \gamma(t) =: \frac{1}{2} f_{tt}
\\ \nonumber
&& g=g(x) , \ g_{xxx} \ne 0 , \ x:= t - \frac{1}{2\tau y} , \ f=f(t) , \ f_{ttt} \ne 0
\end{eqnarray}
Finally, the metric can be brought to the form
\begin{eqnarray}
\label{metryka_NexNe_minusminus_jednasymetria}
ds^{2} &=& 2\phi^{-2} \left\{ \frac{1}{\tau} (d \eta  d w - d \phi  dt) -   2 e^{2\chi_{0}w} \phi^{2}  \,   
dt^{2} -   2 e^{2\chi_{0}w} (2 \chi_{0}t + B_{0}) \phi^{2}  \, dw dt \ \ \ \ \ \ \right.
\\ \nonumber
&& \ \ \ \ \ \ \ \ \ \ \left.
  +  \left( 2 e^{2\chi_{0}w} \eta (\eta +\phi (2\chi_{0}t+B_{0})) - \frac{\phi}{2\tau} (f_{t} + g_{x})  \right) \, dw^{2} \right\} \ \ \ \ \ \ 
\end{eqnarray}
One should mention also that as long as $\chi_{0} \ne 0$, the constant $B_{0}$ can be gauged away.

\subsection{Two symmetries}

As a starting point we take the results from subsection \ref{Sekcja_klasa_I_one_symmetry} written for $\chi_{0}=0$. The first Killing vector reads then $K_{1} = \partial_{w}$. The key function has the form (\ref{funkcja_kluczowa_klasa_I}) with $A=1$, $B=B_{0} = \textrm{const}$, $C=m=n=0$, $F=H(\phi,t)$ where
\begin{equation}
\label{upprossszczone_H}
H_{\phi} = \phi^{2} (g + f)  - \frac{f_{t}}{2 \tau} \phi , \ g=g(x) , \ g_{xxx} \ne 0 , \ x:= t- \frac{1}{2 \tau \phi} , \ f=f(t) , \ f_{ttt} \ne 0
\end{equation}
The metric takes the form
\begin{eqnarray}
\label{metryka_NexNe_minusminus_jednasymetria_izometria}
ds^{2} &=& 2\phi^{-2} \left\{ \frac{1}{\tau} (d \eta  d w - d \phi  dt) -   2  \phi^{2}  \,   
dt^{2} -   2   B_{0} \phi^{2}  \, dw dt \ \ \ \ \ \ \right.
\\ \nonumber
&& \ \ \ \ \ \ \ \ \ \ \left.
  +  \left( 2  \eta (\eta +\phi B_{0}) - \frac{\phi}{2\tau} (f_{t} + g_{x})  \right) \, dw^{2} \right\} \ \ \ \ \ \ 
\end{eqnarray}
Eqs. (\ref{rownanie_master_dla_[N]ex[N]e_minus_minus_1}-\ref{rownanie_master_dla_[N]ex[N]e_minus_minus_5}) give
\begin{equation}
a=a(w) , \ b(w,t) = \chi_{0} t - \frac{1}{2}B_{0} (a-\chi_{0}w) + b_{0} , \ \epsilon(w) =  - \frac{1}{4 \tau^{2}} a_{ww}
\end{equation}
where $b_{0}$ is a constant. Eq. (\ref{rownanie_master_dla_[N]ex[N]e_minus_minus_6}) reads
\begin{eqnarray}
\label{ostateczne_rownanie_dla_[N]ex[N]e_minus_minus}
&& b H_{t} - \chi_{0} \phi H_{\phi} + B_{0} b_{w} \phi^{3} - \tau B_{0} \epsilon \phi^{2} + (\chi_{0} + 2 a_{w}) H 
\\ \nonumber
&& \ \ \ \ \ \
   - \alpha (w,t) \phi^{3} + \frac{1}{2 \tau} b_{ww} \phi^{2} - \frac{1}{2} \epsilon_{w} \phi - \beta (w,t)= 0
\end{eqnarray}
Inserting (\ref{upprossszczone_H}) into the $\partial_{\phi} (\ref{ostateczne_rownanie_dla_[N]ex[N]e_minus_minus})$, after some algebraic work one arrives at the condition $\chi_{0} a_{ww} = 0$, so the cases $\chi_{0} = 0$ and $\chi_{0} \ne 0$ must be considered separately. 

\subsubsection{Case $\chi_{0} \ne 0$}

In the case of the proper homothetic vector we obtain $a=w$, $b=\chi_{0}t$, $\epsilon =0$, $B_{0} (\chi_{0}-1)=0$. The function $g(x)$ reads
\begin{equation}
\label{rozwiazania_na_g_pierwsze}
 g(x) = \left\{
   \begin{array}{ll}
   -g_{0} \ln x - 2 \mu_{0} x + n_{0}  & \textrm{for } \chi_{0} = 1
   \\ 
   -2 \gamma_{0} (x \ln x - x) + m_{0}x + n_{0} & \textrm{for } \chi_{0} = 2
   \\
   \dfrac{ g_{0}\chi_{0}^{2}}{2(\chi_{0}-1)(\chi_{0}-2)} x^{\frac{2\chi_{0}-2}{\chi_{0}}} - 2 \mu_{0}x + n_{0} & \textrm{for } \chi_{0} \ne \{ 1,2 \}
  \end{array}   \right.
\end{equation}
and $f(t)$ has the form
\begin{equation}
\label{rozwiazania_na_f_pierwsze}
 \frac{1}{2} f(t) = \left\{
   \begin{array}{ll}
   -\gamma_{0} \ln t + \mu_{0} t + \nu_{0}  & \textrm{for } \chi_{0} = 1
   \\ 
   \gamma_{0} (t \ln t - t) + \mu_{0}t + \nu_{0} & \textrm{for } \chi_{0} = 2
   \\
   \dfrac{ \gamma_{0}\chi_{0}^{2}}{2(\chi_{0}-1)(\chi_{0}-2)} t^{\frac{2\chi_{0}-2}{\chi_{0}}} + \mu_{0}t + \nu_{0} & \textrm{for } \chi_{0} \ne \{ 1,2 \}
  \end{array}   \right.
\end{equation}
where $\gamma_{0} \ne 0$, $g_{0} \ne 0$, $\mu_{0}$, $\nu_{0}$, $m_{0}$ and $n_{0}$ are constants. Proper homothetic vector $K_{2}$ reads
\begin{equation}
K_{2} = w \partial_{w} + \chi_{0} t \partial_{t} - \chi_{0} \phi \partial_{\phi} - \eta \partial_{\eta}
\end{equation}
In order to get the corresponding metric it is enough to insert (\ref{rozwiazania_na_g_pierwsze}) and (\ref{rozwiazania_na_f_pierwsze}) into (\ref{metryka_NexNe_minusminus_jednasymetria_izometria}).

\subsubsection{Case $\chi_{0} = 0$}

If the second vector is a Killing vector then we get
\begin{equation}
b=-\frac{2}{c_{0}} a_{w} , \ \epsilon = -\frac{1}{4 \tau^{2}} a_{ww} , \ 
a(w) = \left\{
   \begin{array}{ll}
   e^{\frac{B_{0}c_{0}w}{4}} + \dfrac{2b_{0}}{B_{0}}  & \textrm{for } B_{0} \ne 0
   \\ 
   w & \textrm{for } B_{0} = 0
  \end{array}   \right.
\end{equation}
Functions $g(x)$ and $f(t)$ read now
\begin{eqnarray}
\label{rozwiazania_na_gif_drugie}
&& g(x) = \frac{g_{0}}{c_{0}^{3}} e^{c_{0}x} - \dfrac{B_{0}^{2}c_{0}^{2}}{64} x^{2} -2 m_{0} x + n_{0} 
\\ \nonumber
&& \frac{1}{2} f(t) = \frac{\gamma_{0}}{c_{0}^{2}} e^{c_{0}t} + \dfrac{B_{0}^{2}c_{0}^{2}}{128} t^{2} + m_{0} x + s_{0} 
\end{eqnarray}
where $\gamma_{0} \ne 0$, $g_{0} \ne 0$, $c_{0} \ne 0$, $m_{0}$, $n_{0}$, $s_{0}$ and $b_{0}$ are constants. The Killing vector $K_{2}$ has the form
\begin{equation}
K_2 = a \partial_{w}  - \frac{2a_{w}}{c_{0}} \partial_{t} + \left( -a_{w} \eta - \frac{2}{c_{0}} a_{ww} \phi + \frac{1}{4 \tau} a_{ww} \right) \partial_{\eta}
\end{equation}
To obtain the metric one puts (\ref{rozwiazania_na_gif_drugie}) into (\ref{metryka_NexNe_minusminus_jednasymetria_izometria}).

\renewcommand{\arraystretch}{2}


\setcounter{equation}{0}
\section{Type $\{ [\textrm{N}]^{e} \otimes [\textrm{N}]^{n},[--] \}$}
\label{Sekcja_klasa_I_b}

\subsection{Generic case; no symmetries are assumed}

The type  $\{ [\textrm{N}]^{e} \otimes [\textrm{N}]^{n},[--] \}$ does not posses any Lorentzian slice. It is so because of the different properties of the congruences of SD and ASD null strings. The key function for this type has the form (\ref{funkcja_kluczowa_klasa_I}) with $A=0$. It generates the metric (\ref{metryka_typu_NexNe_minusminus_ogolna}) with $A=0$. The field equation (\ref{rownania_pola_klasa_I_row_1}) is an identity. From the field equations (\ref{rownania_pola_klasa_I_row_2}-\ref{rownania_pola_klasa_I_row_3}) and from the transformation formulas for the functions $B,C,m$ and $n$ one finds the solution
\begin{equation}
\label{solution_for_type_NexNn_minusminus_bezsymetrii}
C=m=n=0, \ B=B_{0}=\textrm{const}=  \{ 0,1 \}
\end{equation}
Field equation (\ref{rownania_pola_klasa_I_row_4}) reduces to the form
\begin{equation}
F_{\phi t} + \frac{\gamma}{\tau} \phi = 0
\end{equation}
with the solution
\begin{eqnarray}
\label{solution_for_type_NexNn_minusminus_bezsymetrii_funkcjaF}
&& F_{\phi} = -f \phi + g , \ \ \gamma = \tau f_{t}
\\ \nonumber
&& f=f(w,t), \ f_{tt} \ne 0 , \ g=g(\phi,w),  \ g_{\phi \phi \phi} \ne 0
\end{eqnarray}
Inserting $A=0$, (\ref{solution_for_type_NexNn_minusminus_bezsymetrii}) and (\ref{solution_for_type_NexNn_minusminus_bezsymetrii_funkcjaF}) into (\ref{metryka_typu_NexNe_minusminus_ogolna}) one arrives at the metric
\begin{equation}
\label{metric_for_type_NexNn_minusminus_bezsymetrii}
ds^{2} = 2 \phi^{-2} \left\{ \frac{1}{\tau} (d \eta dw - d \phi dt) - 2B_{0} \phi^{2} dw dt + (2g-\phi g_{\phi} - f \phi + 2B_{0} \eta \phi) dw^{2} \right\}
\end{equation}

\textbf{Remark.} Analysis of Eqs. (\ref{rownanie_master_dla_[N]ex[N]e_minus_minus}) with $A=0$ and with (\ref{solution_for_type_NexNn_minusminus_bezsymetrii}) proves that the metric (\ref{metric_for_type_NexNn_minusminus_bezsymetrii}) with $B_{0}=0$ admits the null Killing vector $K_{0} = \partial_{\eta}$ (see \cite{Chudecki_null_Killing}).

\subsection{One symmetry}
\label{subsection_NexNn_minusminus_onesymmetry}

Symmetries are determined by the Eqs. (\ref{rownanie_master_dla_[N]ex[N]e_minus_minus}) with $A=0$. Homothetic vector can be brought to the form with $a=1$, $b=\epsilon=0$, i.e.
\begin{equation}
K_{1} = \partial_{w} - 2\chi_{0} \left( \phi \partial_{\phi} + \eta \partial_{\eta} \right)
\end{equation}
From the integrability condition (\ref{warunek_calkowalnosci}) we find $\gamma=\gamma(t)$. Eqs. (\ref{rownanie_master_dla_[N]ex[N]e_minus_minus_2}-\ref{rownanie_master_dla_[N]ex[N]e_minus_minus_5}) and (\ref{rownania_pola_klasa_I_row_2}-\ref{rownania_pola_klasa_I_row_3}) can be easily solved and with the help of gauge freedom still available at our disposal one gets the solutions
\begin{equation}
C=m=0 , \ B(w) = B_{0} e^{2 \chi_{0} w} , \ B_{0} = \{0,1 \} , \ n=n_{0} = \textrm{const}
\end{equation}
Eq. (\ref{rownanie_master_dla_[N]ex[N]e_minus_minus_6}) gives the solution for $F$
\begin{equation}
\label{postac_f_NexNn_minusminus}
F(\phi, w, t) = e^{-4\chi_{0}w} H (x,t) + \tilde{\alpha} (w,t) \phi^{3}  , \ x: = \phi e^{2\chi_{0}w}
\end{equation}
where $\tilde{\alpha}$ is an arbitrary function. The last equation which remains to be solved is the field equation (\ref{rownania_pola_klasa_I_row_4}). It reads
\begin{equation}
\label{ostatnie_roww_field_NexNn}
 H_{xt} + \frac{\gamma(t)}{\tau} x
-\tau n_{0}^{2} x - 2\tau n_{0}B_{0} x^{2} - 2  \chi_{0} B_{0} x^{2}  + 3 \tilde{\alpha}_{t}  x^{2} e^{-2\chi_{0}w} = 0
\end{equation}
Differentiating (\ref{ostatnie_roww_field_NexNn}) with respect to $w$ we find the form of the $\tilde{\alpha}$ function, namely $\tilde{\alpha} (w,t) = f_{1}(t) e^{2\chi_{0}w} + f_{2}(w)$, but $f_{2}$ can be gauged away (by using gauge function $L(w)$) and $f_{1}$ can be absorbed into $H$. This proves that $\tilde{\alpha}$ can be put zero without any loss of generality. Hence the solution of Eq. (\ref{ostatnie_roww_field_NexNn}) reads
\begin{equation}
H(x,t) =  \frac{2}{3}B_{0}(\chi_{0} + \tau n_{0}) t x^{3} + \frac{1}{2} (\tau n_{0}^{2} t - f(t)) x^{2} + r(x) + s(t)  , \ \gamma(t) =: \tau f_{t}
\end{equation}
The functions $s(t)$ and $r(x)$ are an arbitrary functions, but $s(t)$ can be gauged away by using the gauge function $M$. Denoting $r_{x} =: g(x)$ we arrive at the metric
\begin{eqnarray}
\label{metryka_NexNn_minusminus_jednasymetria}
ds^{2} &=& 2\phi^{-2} \left\{ \frac{1}{\tau} (d \eta  d w - d \phi  dt) - 2 B_{0} e^{2\chi_{0}w}  \phi^{2}  \, dw dt \ \ \ \ \ \ \right.
\\ \nonumber
&& \ \ \ \ \ \ \ \ \ \ \left.
  +  \left( 2 B_{0} x \eta  + 2 n_{0} \eta + e^{-2\chi_{0}w} \left( (\tau n_{0}^{2}t-f)x +2g-xg_{x} \right)  \right) \, dw^{2} \right\} \ \ \ \ \ \ 
\end{eqnarray}

The metric (\ref{metryka_NexNn_minusminus_jednasymetria}) can be equipped with additional, null Killing vector. Such a case requires $B_{0}=0$ and the null Killing vector reads then $K_{0} = e^{-2 \tau n_{0}w} \partial_{\eta}$.

\subsection{Two symmetries}

First, all results from subsection \ref{subsection_NexNn_minusminus_onesymmetry} we can adapt to the case of the Killing vector $K_{1} = \partial_{w}$. Inserting $\chi_{0}=0$ into the results from subsection \ref{subsection_NexNn_minusminus_onesymmetry} we find 
\begin{eqnarray}
&& C=m=0 , \ B = B_{0}  = \{0,1 \} , \ n=n_{0} = \textrm{const}
\\ \nonumber
&& F(\phi, t) =  \frac{2}{3}B_{0} \tau n_{0} t \phi^{3} + \frac{1}{2} (\tau n_{0}^{2} t - f(t)) \phi^{2} + r(\phi)   , \ \gamma(t) =: \tau f_{t}
\end{eqnarray}
what generates the metric
\begin{eqnarray}
\label{metryka_NexNn_minusminus_jednasymetria_izometria}
ds^{2} &=& 2\phi^{-2} \left\{ \frac{1}{\tau} (d \eta  d w - d \phi  dt) - 2 B_{0}   \phi^{2}  \, dw dt \ \ \ \ \ \ \right.
\\ \nonumber
&& \ \ \ \ \ \ \ \ \ \ \left.
  +  \left( 2 B_{0} \phi \eta  + 2 n_{0} \eta +  \left( (\tau n_{0}^{2}t-f)\phi +2g-\phi g_{\phi} \right)  \right) \, dw^{2} \right\} \ \ \ \ \ \ 
\end{eqnarray}
with $g(\phi) := r_{\phi}$. 

Eqs. (\ref{rownanie_master_dla_[N]ex[N]e_minus_minus}) read now 
\begin{eqnarray}
\label{symetrie_drugie_NexNn}
&& b_{tt} = 0 , \ B_{0} (b_{t} + a_{w} - 2\chi_{0} ) , \ 2 \tau n_{0} a_{w} - a_{ww} + 2b_{tw} = 0 , \ \epsilon = \epsilon (w)
\\ \nonumber
&& bF_{t} + (b_{t} - 2 \chi_{0}) \phi F_{\phi} + (4 \chi_{0} + 2a_{w} -3b_{t}) F
+ b_{w} (B_{0} \phi^{3} + n_{0} \phi^{2}) - \tau \epsilon (B_{0} \phi^{2} + n_{0} \phi )
\\ \nonumber
&& \ \ \ \ \ \ \ \ \ \ \ \ \ \ 
-\alpha \phi^{3} + \frac{1}{2\tau} b_{ww} \phi^{2} - \frac{1}{2} \epsilon_{w} \phi  - \beta = 0
\end{eqnarray}
To find the second homothetic vector one has to solve Eqs. (\ref{symetrie_drugie_NexNn}). It must be remembered, that all subcases with $B_{0}=0$ admit additional, null Killing vector $K_{0} = e^{-2 \tau n_{0}w} \partial_{\eta}$. 

The great number of subcases does not allow us to present any details. We give only final results. All metrics of the spaces of the types $\{ [\textrm{N}]^{e} \otimes [\textrm{N}]^{n},[--] \}$ equipped with at least two symmetries can be obtained from (\ref{metryka_NexNn_minusminus_jednasymetria_izometria}). It is enough to put into (\ref{metryka_NexNn_minusminus_jednasymetria_izometria}) functions $f$ and $g$ in the forms given below.

\subsubsection{Case $B_{0} \ne 0$, $\chi_{0} \ne 0$}

We obtain $a(w) = w$, $b (w,t) = (2 \chi_{0} -1)t + b_{0}$, if $2 \chi_{0} =1$ then necessarily  $b_{0} \ne 0$, $\epsilon = \epsilon_{0} = \textrm{const}$ and $n_{0} = 0$; $\epsilon_{0}$ reads
\begin{equation}
 \epsilon_{0} = \left\{
   \begin{array}{ll}
    \dfrac{\tau g_{0} - \gamma_{0}}{2 \tau^{2} B_{0}}  & \textrm{for } \chi_{0} = \dfrac{3}{2}
   \\
   \dfrac{(2\chi_{0}-3)(r_{0}-\tau \nu_{0})}{2 \tau^{2} B_{0}}  & \textrm{for } \chi_{0} \ne \dfrac{3}{2}
   \\ 
  \end{array}   \right.
\end{equation}
and
\begin{equation}
 g(\phi) = \left\{
   \begin{array}{ll}
   g_{0} \left( \frac{1}{2} \phi^{2} \ln \phi - \frac{3}{4} \phi^{2} \right) + \frac{1}{2} \mu_{0} \phi^{2} + \nu_{0} \phi   & \textrm{for } \chi_{0} = 1
   \\ 
   -g_{0} (  \phi \ln \phi -  \phi ) + \frac{1}{2} \mu_{0} \phi^{2} + \nu_{0} \phi  & \textrm{for } \chi_{0} = \dfrac{3}{2}
   \\
   \dfrac{g_{0} \phi^{4-2\chi_{0}}}{4(1-\chi_{0})(3-2\chi_{0})(2-\chi_{0})} + \frac{1}{2} \mu_{0} \phi^{2} + \nu_{0} \phi & \textrm{for } \chi_{0} \ne \left\{ 1, \dfrac{3}{2}, 2 \right\}
  \end{array}   \right.
\end{equation}
\begin{equation}
 \tau f(t) = \left\{
   \begin{array}{ll}
   - \dfrac{\gamma_{0}b_{0}}{2} e^{- \frac{2t}{b_{0}}} + r_{0}  & \textrm{for } \chi_{0} = \dfrac{1}{2}
   \\ 
  \dfrac{\gamma_{0}}{2} \ln (2t+b_{0}) +r_{0} & \textrm{for } \chi_{0} = \dfrac{3}{2}
   \\
   \dfrac{\gamma_{0}}{2\chi_{0}-3} ((2 \chi_{0}-1)t+b_{0})^{\frac{2 \chi_{0}-3}{2 \chi_{0}-1}} +r_{0} & \textrm{for } \chi_{0} \ne \left\{ \dfrac{1}{2}, \dfrac{3}{2} \right\}
  \end{array}   \right.
\end{equation}
where $g_{0} \ne 0$, $\gamma_{0} \ne 0$, $\mu_{0}$, $\nu_{0}$ and $r_{0}$ are constants. 

\textbf{Remark.} Extremely interesting behavior of the ASD conformal curvature can be observed in this subcase. For $\chi_{0} = 2$ one gets $g_{\phi \phi \phi}=0$, what causes that ASD conformal curvature vanishes and the space automatically reduces to the right-conformally flat space.

\subsubsection{Case $B_{0} \ne 0$, $\chi_{0} = 0$}

Here we have
\begin{equation}
b= -a_{w} t , \ \epsilon = \dfrac{3a_{w} (s_{0} - d_{0})}{2 \tau B_{0}} , \ 
a(w) = \left\{
   \begin{array}{ll}
   a_{0} e^{\frac{2}{3} \tau n_{0} w} + \textrm{const}  & \textrm{for } n_{0} \ne 0
   \\ 
   w & \textrm{for } n_{0} = 0
  \end{array}   \right.
\end{equation}
and
\begin{eqnarray}
&& g( \phi ) = \frac{g_{0}}{24} \phi^{4} + \frac{r_{0}}{2} \phi^{2} + s_{0} \phi + \frac{n_{0}(s_{0}-d_{0})}{2  B_{0}}
\\ \nonumber
&& f(t) = c_{0} t^{3} + \frac{\tau n_{0}^{2}}{9} t + d_{0}
\end{eqnarray}
where $a_{0} \ne 0$, $g_{0} \ne 0$, $c_{0} \ne 0$, $r_{0}$, $s_{0}$ and $d_{0}$ are constants.

\subsubsection{Case $B_{0} = 0$, $\chi_{0} \ne 0$}

The constant $n_{0}$ must vanish in this case, $n_{0}=0$. Moreover $a(w)=w$, $b(w,t) = c_{0} t + b_{0}$, but if $c_{0} = 0$, then necessarily $b_{0} \ne 0$, and if $c_{0} \ne 0$, then $b_{0}=0$. We also get
\begin{equation}
 \epsilon (w) = \left\{
   \begin{array}{ll}
    4s_{0} (\chi_{0}-1) w + \epsilon_{0}  & \textrm{for } c_{0}=0, \chi_{0} \ne 1
   \\
   4s_{0} (2-c_{0}) w + \epsilon_{0}  & \textrm{for } c_{0} \ne 2, \chi_{0} = 1
   \\
   g_{0} (2-c_{0}) w + \epsilon_{0}  & \textrm{for } c_{0} \ne 2, \chi_{0} \ne 1, \chi_{0} = c_{0} -1
   \\
   4s_{0} (\chi_{0}+1-c_{0}) w + \epsilon_{0}  & \textrm{for } c_{0} \ne 2, \chi_{0} \ne 1, \chi_{0} \ne c_{0} -1
  \end{array}   \right.
\end{equation}
\begin{equation}
 g(\phi) = \left\{
   \begin{array}{ll}
   g_{0} \left( \frac{1}{2} \phi^{2} \ln \phi - \frac{3}{4} \phi^{2} \right) + \frac{1}{2} \mu_{0} \phi^{2} + \dfrac{f_{0}}{\tau} \phi  + s_{0} & \textrm{for } c_{0} \ne 2, \chi_{0} = 1
   \\ 
  -g_{0} (\phi \ln \phi- \phi) + \frac{1}{2} \mu_{0} \phi^{2} + r_{0} \phi + s_{0} & \textrm{for } c_{0} = 2, \chi_{0} \ne 1
  \\
  \dfrac{g_{0}}{2} \ln \phi + \frac{1}{2} \mu_{0} \phi^{2} +  \dfrac{f_{0}}{\tau}  \phi + s_{0} & \textrm{for } c_{0} \ne 2, \chi_{0} \ne 1, \chi_{0} = c_{0}-1
  \\
  \dfrac{g_{0} (2\chi_{0}-c_{0})^{3}}{4(1-\chi_{0})(2-c_{0})(\chi_{0}-c_{0}+1)} \phi^{\frac{2(\chi_{0}-c_{0}+1)}{2\chi_{0}-c_{0}}} & 
  \\
  \ \ \ \ \ \ \ \ \ \ \ \ \ \ \ \ 
+ \frac{1}{2} \mu_{0} \phi^{2} +  \dfrac{f_{0}}{\tau}  \phi + s_{0} & \textrm{for } c_{0} \ne 2, \chi_{0} \ne 1, \chi_{0}  \ne c_{0}-1
  \end{array}   \right.
\end{equation}
\begin{equation}
 \tau f(t) = \left\{
   \begin{array}{ll}
   - \dfrac{\gamma_{0} b_{0}}{2} e^{- \frac{2t}{b_{0}}} + f_{0} & \textrm{for } c_{0} =0
   \\ 
   \tau g_{0} (\chi_{0}-1) \ln (2t + b_{0}) + f_{0}  & \textrm{for } c_{0} =2
   \\
   \dfrac{\gamma_{0}}{c_{0}-2} (c_{0}t + b_{0})^{\frac{c_{0}-2}{c_{0}}} + f_{0} &  \textrm{for } c_{0} \ne \{0,2  \}
  \end{array}   \right.
\end{equation}
where $g_{0} \ne 0$, $\gamma_{0} \ne 0$, $\mu_{0}$, $f_{0}$ and $s_{0}$ are constants.

\subsubsection{Case $B_{0} = 0$, $\chi_{0} = 0$, $n_{0} \ne 0$}

In this case
\begin{equation}
a(w) = \frac{a_{0}}{c_{0}} e^{c_{0}w} + e_{0}, \ b(w,t) = \left( \frac{1}{2} - \frac{ \tau n_{0}}{c_{0}} \right) a_{w} t
\end{equation}
and the constant $c_{0}$ must satisfy the constraint $2 \tau n_{0} \ne c_{0}$. Moreover
\begin{equation}
\epsilon (w) = \left\{
   \begin{array}{ll}
  \left( \dfrac{2s_{0}a_{0}}{c_{0}} + \epsilon_{0}e^{2c_{0}w}  \right)  e^{c_{0}w} & \textrm{for } 2 \tau n_{0} = -3 c_{0}
  \\
  (a_{0}g_{0}w + \epsilon_{0}) e^{c_{0}w} & \textrm{for } 2 \tau n_{0} = - c_{0}
  \\
  \left( \dfrac{2s_{0}a_{0}}{c_{0}} + \epsilon_{0}e^{-(c_{0}+2 \tau n_{0})w}  \right)  e^{c_{0}w} & \textrm{for } 2 \tau n_{0} \ne \{ -c_{0}, -3c_{0} \}
  \end{array}   \right.
\end{equation}
\begin{equation}
 g(\phi) = \left\{
   \begin{array}{ll}
  \dfrac{\gamma_{0}}{\tau} (\phi \ln \phi- \phi) + \frac{1}{2} \mu_{0} \phi^{2} + r_{0} \phi + s_{0} & \textrm{for } 2 \tau n_{0} = -3 c_{0}
  \\
  \dfrac{g_{0}}{2} \ln \phi + \frac{1}{2} \mu_{0} \phi^{2} +  \dfrac{f_{0}}{\tau}  \phi + s_{0} & \textrm{for } 2 \tau n_{0} = - c_{0}
  \\
  - \dfrac{g_{0} (c_{0}-2 \tau n_{0})^{3}}{8c_{0}(3c_{0}+2 \tau n_{0})(c_{0}+2 \tau n_{0})} \phi^{\frac{2(2 \tau n_{0} + c_{0})}{2 \tau n_{0} - c_{0}}} & 
  \\
  \ \ \ \ \ \ \ \ \ \ \ \ \ \ \ \ 
+ \frac{1}{2} \mu_{0} \phi^{2} +  \dfrac{f_{0}}{\tau}  \phi + s_{0} & \textrm{for } 2 \tau n_{0} \ne \{ -c_{0}, -3c_{0} \}
  \end{array}   \right.
\end{equation}
\begin{equation}
 \tau f(t) = \left\{
   \begin{array}{ll}
  \gamma_{0} \ln t + \dfrac{c_{0}^{2}}{4} t + f_{0} & \textrm{for } 2 \tau n_{0} = -3 c_{0}
  \\
  \dfrac{\gamma_{0} (2 \tau n_{0} - c_{0})}{2 \tau n_{0} + 3c_{0}} t^{\frac{2 \tau n_{0} +3 c_{0}}{2 \tau n_{0} - c_{0}}} + \dfrac{c_{0}^{2}}{4} t + f_{0} & \textrm{for } 2 \tau n_{0} \ne  -3 c_{0}
  \end{array}   \right.
\end{equation}
where $a_{0} \ne 0$, $c_{0} \ne 0$, $g_{0} \ne 0$, $\gamma_{0} \ne 0$, $\mu_{0}$, $r_{0}$, $s_{0}$, $f_{0}$ and  $\epsilon_{0}$ are constants.

\subsubsection{Case $B_{0} = 0$, $\chi_{0} = 0$, $n_{0} = 0$, $a_{ww} = 0$}

In this case we find $a(w)=w$, $b (t) = \frac{1}{2}(b_{0}+1) t$, where $b_{0}$ is a constant such that $b_{0} \ne -1$. Function $\epsilon$ reads
\begin{equation}
\epsilon (w) = \left\{
   \begin{array}{ll}
  - 4s_{0}w+\epsilon_{0} & \textrm{for } b_{0}=3
  \\
  g_{0}w + \epsilon_{0} & \textrm{for } b_{0}=1
  \\
  2(1-b_{0})s_{0}w+ \epsilon_{0} & \textrm{for } b_{0} \ne \{ 1,3 \}
  \end{array}   \right.
\end{equation}
Then
\begin{equation}
 g(\phi) = \left\{
   \begin{array}{ll}
  \dfrac{\gamma_{0}}{\tau} (\phi \ln \phi- \phi) + \frac{1}{2} \mu_{0} \phi^{2} + r_{0} \phi + s_{0} & \textrm{for } b_{0} = 3
  \\
  \dfrac{g_{0}}{2} \ln \phi + \frac{1}{2} \mu_{0} \phi^{2} +  \dfrac{f_{0}}{\tau}  \phi + s_{0} & \textrm{for } b_{0} = 1
  \\
  - \dfrac{g_{0} (b_{0}+1)^{3}}{8(b_{0}-3)(b_{0}-1)} \phi^{\frac{2(b_{0}-1)}{b_{0}+1}} & 
  \\
  \ \ \ \ \ \ \ \ \ \ \ \ \ \ \ \ 
+ \frac{1}{2} \mu_{0} \phi^{2} +  \dfrac{f_{0}}{\tau}  \phi + s_{0} & \textrm{for } b_{0}  \ne \{ 1,3 \}
  \end{array}   \right.
\end{equation}
\begin{equation}
 \tau f(t) = \left\{
   \begin{array}{ll}
  \gamma_{0} \ln t + f_{0} & \textrm{for } b_{0} =3
  \\
  \dfrac{\gamma_{0} (b_{0} +1)}{b_{0}-3} t^{\frac{b_{0}-3}{b_{0}+1}}  + f_{0} & \textrm{for } b_{0} \ne 3
  \end{array}   \right.
\end{equation}
where $g_{0} \ne 0$, $\gamma_{0} \ne 0$, $\mu_{0}$, $r_{0}$, $s_{0}$, $\epsilon_{0}$ and $f_{0}$ are constants.

\subsubsection{Case $B_{0} = 0$, $\chi_{0} = 0$, $n_{0} = 0$, $a_{ww} \ne 0$}

Last case is characterized by $b (w,t) = \frac{1}{2} a_{w} t$, $\epsilon (w) = 2s_{0} a +\epsilon_{0}$ and
\begin{eqnarray}
&& g(\phi) = - \dfrac{g_{0}}{24} \phi^{-2} + \frac{1}{2} \mu_{0} \phi^{2} + \dfrac{f_{0}}{2} \phi + s_{0}
\\ \nonumber
&& \tau f(t) = - \dfrac{\gamma_{0}}{3}  t^{-3} + \dfrac{c_{0}}{4}t + f_{0}
\end{eqnarray}
Then $a(w)$ is an arbitrary function which satisfies the equation $a_{www} = c_{0} a_{w}$. Moreover, $g_{0} \ne 0$, $\gamma_{0} \ne 0$, $c_{0}$, $f_{0}$, $\mu_{0}$, $r_{0}$ i $s_{0}$ are constants.


\setcounter{equation}{0}
\section{Type $\{ [\textrm{N}]^{e} \otimes [\textrm{N}]^{e},[+-] \}$}
\label{Sekcja_klasa_II}

\subsection{Generic case; no symmetries are assumed}

The key function is given by the formula (\ref{funkcja_kluczowa_dla_NexNe_plusminus}) and it generates the metric in the form
\begin{eqnarray}
\label{general_form_metric_NexNe_plusminus}
ds^{2} &=& 2 \phi^{-2} \{ \tau^{-1} (d \eta dw - d \phi dt) - \phi (F_{\eta \eta} + 2B \phi ) dt^{2}
\\ \nonumber
&& \ \ \ \ \ \ \ \ +2F_{\eta} \, dw dt + 2 (B \eta^{2} +C \eta +m ) dw^{2} \}
\end{eqnarray}
Careful analysis of the transformation formulas for the functions $A,B,C$ and $m$ which follow from (\ref{transformacja_funkcji_kluczowej}) shows, that without any loss of generality one can put $A=0$. Further steps depend on the function $B$. Cases $B=0$ and $B \ne 0$ must be considered separately.

\subsubsection{Case $B = 1$}
\label{Subsubsekcja_klasa_II_Bjeden_bez_symetrii}

If $B\ne 0$ then one can put $B=1$. Moreover $A=C=0$ and $m=m(w,t) \ne 0$. From (\ref{rownania_pola_typNexNe_plusminus}) we get
\begin{subequations}
\begin{eqnarray}
&&  \frac{\gamma}{\tau^{2}} =  - 4m , \ m_{t} \ne 0
\\ 
\label{ogolne_rownanie_pola_typ_NexNe_plusminus}
&& 2( \eta^{2}  +m) F_{\eta \eta} - 4 \eta  F_{\eta} - \frac{1}{\tau} F_{\eta w} + \frac{1}{\tau} m_{t}=0 , \ F_{\eta \eta \eta \eta} \ne 0
\end{eqnarray}
\end{subequations}
The general solution of Eq. (\ref{ogolne_rownanie_pola_typ_NexNe_plusminus}) is not known. However, under additional \textsl{ad hoc} assumption $m_{w} = 0 \ \Longleftrightarrow \ m=m(t)$ the solution reads
\begin{eqnarray}
&& F_{\eta} = -\frac{m_{t} (m+ \eta^{2})}{4 \tau m} \left( \frac{\eta}{m+\eta^{2}} - 2 \tau w + f \right)
\\ \nonumber
&& f=f(x,t) , \ x:= w + \frac{1}{2 \tau m^{\frac{1}{2}}} \, \textrm{arctg} \left( \frac{\eta}{m^{\frac{1}{2}}} \right)
\end{eqnarray}

Return now to the case with $m=m(w,t)$. If we change the coordinates and function $F$ according to the formulas
\begin{equation}
\phi =: \frac{1}{r} , \ t =: -\tau u , \ w= :\frac{1}{2} \bar{\zeta} , \ F_{\eta} =: \Sigma (\eta,u, \bar{\zeta})
\end{equation}
then the metric (\ref{general_form_metric_NexNe_plusminus}) and Eq. (\ref{ogolne_rownanie_pola_typ_NexNe_plusminus}) take the form
\begin{subequations}
\label{metryka_i_rownanie_klasa_Rob_Traut_1}
\begin{eqnarray}
&& ds^{2} = -2dr du - (2\tau^{2}r \Sigma_{\eta} + 4 \tau^{2}) du^{2} + r^{2} \left( \frac{1}{\tau} d \eta + (\eta^{2} +m) d \bar{\zeta} - 2 \tau \Sigma \, du \right) d \bar{\zeta} \ \ \ \ \
\\ 
&& 2 (\eta^{2} +m) \Sigma_{\eta} -  4 \eta \Sigma - \frac{2}{\tau} \Sigma_{\bar{\zeta}} - \frac{1}{\tau^{2}} m_{u} = 0
\end{eqnarray}
\end{subequations}
Consider the coordinate transformation $\eta \longrightarrow \zeta$ such that $\eta(u, \zeta, \bar{\zeta}) = \tau^{-1} ( \ln P)_{\bar{\zeta}}$. Function $P=P(u, \zeta, \bar{\zeta})$ is chosen in such a manner that $2 \tau^{3} \, \Sigma (u, \zeta, \bar{\zeta}) =  (\ln P)_{u \bar{\zeta}}$. Formulas (\ref{metryka_i_rownanie_klasa_Rob_Traut_1}) read now
\begin{subequations}
\label{metryka_i_rownanie_klasa_Rob_Traut_2}
\begin{eqnarray}
&& ds^{2} = -2dr du - \big( r \, \partial_{u} \ln ( (\ln P)_{\zeta \bar{\zeta}}  ) + 4 \tau^{2} \big) du^{2} 
\\ \nonumber
&& \ \ \ \ \ \ \ \ \ 
+ r^{2} \left( \frac{1}{\tau^{2}} \big( \tau^{2} m + (\ln P)^{2}_{\bar{\zeta}} + (\ln P)_{\bar{\zeta} \bar{\zeta}} \big) \, d \bar{\zeta} + \frac{1}{\tau^{2}} (\ln P )_{\zeta \bar{\zeta}} \, d \zeta \right) d \bar{\zeta} \ \ \ \ \
\\ 
\label{metryka_i_rownanie_klasa_Rob_Traut_2_row}
&& \tau^{2} m + (\ln P)^{2}_{\bar{\zeta}} + (\ln P)_{\bar{\zeta} \bar{\zeta}} = h( \zeta, \bar{\zeta}) (\ln P)_{\zeta \bar{\zeta}}
\end{eqnarray}
\end{subequations}
In (\ref{metryka_i_rownanie_klasa_Rob_Traut_2_row}) function $h$ is an arbitrary function, but it can be gauged away without any loss of generality. Because $m=m(u, \bar{\zeta})$, differentiation of Eq. (\ref{metryka_i_rownanie_klasa_Rob_Traut_2_row}) with respect to $\zeta$ gives $\partial_{\bar{\zeta}} (2P^{2} (\ln P)_{\zeta \bar{\zeta}})=0$. Choosing the constant $\tau$ in such a manner that $4 \tau^{2} = \varepsilon := \pm 1$ one arrives at the final form of the metric
\begin{equation}
\label{metryka_i_rownanie_klasa_Rob_Traut_3}
 ds^{2} = -2dr du - \big( r \, \partial_{u} \ln ( KP^{-2}  ) + \varepsilon \big) du^{2} + \frac{2r^{2}}{P^{2}} \frac{K}{\varepsilon} \, d\zeta d\bar{\zeta}
\end{equation}
In (\ref{metryka_i_rownanie_klasa_Rob_Traut_3}) functions $P=P(u, \zeta, \bar{\zeta})$ and $K=K(u, \zeta)$ are constrained by the equation 
\begin{equation}
\label{metryka_i_rownanie_klasa_Rob_Traut_3_row}
2P^{2} (\ln P)_{\zeta \bar{\zeta}} = K
\end{equation}
It must be remembered that at this stage the metric (\ref{metryka_i_rownanie_klasa_Rob_Traut_3}) is still the complex metric (all coordinates are complex, all functions are holomorphic, bar does not mean anything).

To get Lorentzian slice of the metric (\ref{metryka_i_rownanie_klasa_Rob_Traut_3}) we have to consider the coordinates $(r,u)$ as real ones, coordinates $(\zeta, \bar{\zeta})$ as complex (where bar denotes the complex conjugation), functions $P$ and $K$ as real analytic. Function $K$ becomes the function of only one variable $u$ and admissible gauge freedom allows to bring it to the constant value, $K = \varepsilon$. The metric (\ref{metryka_i_rownanie_klasa_Rob_Traut_3}) becomes real metric with Lorentzian signature where function $P$ satisfies the equation $2P^{2} (\ln P)_{\zeta \bar{\zeta}} = \pm 1$. This is exactly the general class of the Lorentzian metrics known as the vacuum type $[\textrm{N}]$ Robinson - Trautman solutions (compare \cite{Exacty}, Theorem 28.1 specialized for $m=0$ and $\Delta \ln P=\pm 1$).

\subsubsection{Case $B = 0$}
\label{Subsubsekcja_klasa_II_Bzero_bez_symetrii}

If $B=0$ then $A=m=0$ and $C=C(w,t) \ne 0$. From (\ref{rownania_pola_typNexNe_plusminus}) one finds
\begin{eqnarray}
&& \frac{\gamma}{\tau^{2}} = C^{2} + \frac{1}{\tau} C_{w} 
\\ \nonumber
&& 2C \eta F_{\eta \eta} - 2C F_{\eta} - \frac{1}{\tau} F_{\eta w} + \frac{1}{\tau} C_{t} \eta = 0
\end{eqnarray}
This case can be simply solved and the solution reads
\begin{eqnarray}
\label{rozwiazanie_NexNe_plus_minus_bez_symetrii}
&&  C=:f_{w} , \ f=f(w,t) , \ (\tau f_{w}^{2} + f_{ww})_{t} \ne 0
\\ \nonumber
&& F_{\eta} = \eta (g+f_{t}) , \ g=g(x,t) , \ x:= \eta e^{2 \tau f}, \ (xg)_{xxx} \ne 0
\end{eqnarray}
The metric (\ref{general_form_metric_NexNe_plusminus}) written in coordinates $(\phi,x,w,t)$ reads
\begin{equation}
\label{general_form_metric_NexNe_plusminus_step0}
ds^{2} = 2 \phi^{-2} \left\{ \frac{1}{\tau} e^{-2\tau f} \, dwdx - \frac{1}{\tau} \, d \phi dt - \phi (f_{t} + (xg)_{x} ) \, dt^{2} + 2e^{-2\tau f} xg \, dw dt \right\}
\end{equation}
To find Lorentzian slice we change coordinates $(\phi,t,w)$ and the function $f$ according to the formulas
\begin{equation}
\label{changing_coordinates_first_step}
\phi =: \frac{1}{r} , \ t=:-\tau u, \ w =: \frac{1}{2} \bar{\zeta} , \ f(w,t) =: \frac{1}{\tau} \ln \bar{H} , \ \bar{H} = \bar{H} (u, \bar{\zeta}) , \ \tau^{2} \, xg =: G(x,u)
\end{equation}
Feeding (\ref{general_form_metric_NexNe_plusminus_step0}) with (\ref{changing_coordinates_first_step}) one arrives at the form
\begin{equation}
\label{general_form_metric_NexNe_plusminus_step1}
ds^{2} = -2drdu + 2r ( \partial_{u} \ln \bar{H} - G_{x}) \, du^{2} + \frac{r^{2}}{\tau \bar{H}^{2}} \, (dx-2G\, du) d \bar{\zeta}
\end{equation}
Then we note that there exist functions $h=h(x,u)$ and $\zeta = \zeta (x,u)$ such that 
\begin{equation}
\frac{1}{2\tau} \, (dx-2G\, du) = \frac{1}{h^{2}} \, d\zeta
\end{equation}
(it is so, because in two dimensions each vector is proportional to a gradient). Treating now $\zeta$ as an independent variable one finds $G_{x} = - \partial_{u} \ln H$ where $H(u, \zeta) := h(x(\zeta, u),u)$. Finally we obtain the metric in the form
\begin{equation}
\label{general_form_metric_NexNe_plusminus_step2}
ds^{2} = -2drdu + 2r \, \partial_{u} \ln (H\bar{H})  \, du^{2} + \frac{2r^{2}}{H^{2} \bar{H}^{2}} \, d \zeta d \bar{\zeta}
\end{equation}
Consider now the coordinates $(r,u)$ as real ones and the coordinates $(\zeta, \bar{\zeta})$ as complex (where bar denotes the complex conjugation). The metric (\ref{general_form_metric_NexNe_plusminus_step2}) becomes the real metric with Lorentzian signature. This is exactly a special class of the vacuum type $[\textrm{N}]$ Robinson - Trautman metrics (compare \cite{Exacty}, Theorem 28.1 specialized for $m=0$ and $\Delta \ln P=0$). Hence
\begin{eqnarray}
\nonumber
 \{ [\textrm{N}]^{e} \otimes [\textrm{N}]^{e},[+-] \} \textrm{ with } B \ne 0 & \stackrel{\textrm{real Lorentzian slice}}{\longrightarrow} & \textrm{Robinson - Trautman solution}
 \\ \nonumber
 && \textrm{with } \Delta \ln P=\pm 1
\\ \nonumber
 \{ [\textrm{N}]^{e} \otimes [\textrm{N}]^{e},[+-] \} \textrm{ with } B = 0 & \stackrel{\textrm{real Lorentzian slice}}{\longrightarrow} & \textrm{Robinson - Trautman solution}
\\ \nonumber
&& \textrm{with } \Delta \ln P=0
\end{eqnarray}

\subsection{One symmetry}
\label{jedna_symetria_NexNe_plusminus}

Putting the key function (\ref{funkcja_kluczowa_dla_NexNe_plusminus}) into the master equation (\ref{master_equation}) the relations between functions $F, A, B, C, m$ and $a,b, \epsilon, \alpha, \beta$ can be found. They read
\begin{subequations}
\label{uklad_master_typ_NexNe_plusminus}
\begin{eqnarray}
&& aA_{w} + (2a_{w} - 2\chi_{0}) A - \alpha = 0
\\ 
&& b_{w}=0
\\ 
&& bB_{t} + (2b_{t} - 2\chi_{0}) B= 0
\\ 
&& aC_{w} + bC_{t} + a_{w}C - 2 \tau \epsilon B - \frac{1}{2\tau} a_{ww} = 0
\\ 
&& am_{w} + bm_{t} + (2 \chi_{0} + 2a_{w} -2 b_{t}) m - \tau \epsilon C - \frac{1}{2} \epsilon_{w} = 0
\\ 
&& aF_{w} + bF_{t} + (2 b_{t} - a_{w} -2 \chi_{0}) \eta F_{\eta} - \tau \epsilon F_{\eta} 
\\ \nonumber
&& \ \ \ \ \ \ \ \ \ \ \ \ 
+ (4 \chi_{0}+ 2a_{w} - 3 b_{t}) F + \frac{1}{2 \tau} b_{tt} \eta^{2} - \frac{1}{2} \epsilon_{t} \eta -\beta = 0
\end{eqnarray}
\end{subequations}

\subsubsection{Case $B=1$}
\label{jedna_symetria_B1_NexNe_plusminus}

With $A=C=0$, $B=1$ and $m=m(w,t) \ne 0$ one finds that the proper homothetic vector can be always brought to the form

\begin{equation}
K_{1} = \chi_{0} \left( \partial_{w} + t \partial_{t}  - \phi \partial_{\phi} \right)
\end{equation}
and the solutions of the system (\ref{uklad_master_typ_NexNe_plusminus}) are
\begin{equation}
m = m(x) , \ m_{x} \ne 0 , \ F = t^{-1} H (\eta, x) , \ H_{\eta \eta \eta \eta} \ne 0 , \ x:= w- \ln t
\end{equation}
The field equation (\ref{ogolne_rownanie_pola_typ_NexNe_plusminus}) reduces to the linear second order PDE
\begin{equation}
\label{zredukowane_rownanie_pola_dla_dla_NexNe_plusminus}
2 (\eta^{2} + m) H_{\eta \eta} - 4 \eta H_{\eta} - \frac{1}{\tau} H_{\eta x} - \frac{1}{\tau} m_{x} =0
\end{equation}

For the Killing symmetries we find two possibilities. The first is 
\begin{equation}
K_{1}   = \partial_{w} + \partial_{t}
\end{equation}
Hence
\begin{equation}
\label{pomocnicze_label_na_def_x}
m = m(x) , \ m_{x} \ne 0 , \  F =  H (\eta, x) , \ H_{\eta \eta \eta \eta} \ne 0 , \ x:= w-  t
\end{equation}
and the field equations reduce to the same equation (\ref{zredukowane_rownanie_pola_dla_dla_NexNe_plusminus}) but with $x$ defined by (\ref{pomocnicze_label_na_def_x}).   

The second possibility is the Killing vector in the form
\begin{equation}
K_{1} =  \partial_{w}
\end{equation}
In this case the general solution can be found and it reads
\begin{eqnarray}
\label{funkcja_F_po_eta_dla_NexNe_plusminus}
&& F_{\eta} = -\frac{m_{t} (m+ \eta^{2})}{4 \tau m^{\frac{3}{2}}} \left( \frac{m^{\frac{1}{2}} \eta}{m+\eta^{2}} + \textrm{arctg} \, \left( \frac{\eta}{m^{\frac{1}{2}}} \right)  + g \right) , \ g=g(t) 
\\ \nonumber
&&  m=m(t) , \ m_{t} \ne 0
\end{eqnarray}
\textbf{Remark}: Substitution $H_{\eta} = e^{G} - m - \eta^{2}$ reduces Eq. (\ref{zredukowane_rownanie_pola_dla_dla_NexNe_plusminus}) to the following linear first order PDE
\begin{equation}
2 (m+\eta^{2}) G_{\eta} - \frac{1}{\tau} G_{x} - 4 \eta = 0
\end{equation}
which looks better then the Eq. (\ref{zredukowane_rownanie_pola_dla_dla_NexNe_plusminus}) but still the general solution is unknown.

\subsubsection{Case $B=0$}
\label{subsubsection_Przypadek_Bzero}

Here, analogously as before, we find two possibilities. The first is characterized by $a=1$ and $b =0$. The homothetic vector reads
\begin{equation}
K_{1} = \partial_{w} - 2 \chi_{0} \left( \phi \partial_{\phi} + \eta \partial_{\eta} \right)
\end{equation}
and the solution has the form (\ref{rozwiazanie_NexNe_plus_minus_bez_symetrii}) with 
\begin{equation}
f(w,t) = C w , \ C=C(t) , \ 
g(x,t) = h - \frac{C_{t} \ln x}{2\tau C - 2 \chi_{0}} , \ h=h(t)
\end{equation}
where $C=C(t)$ and $h=h(t)$ are arbitrary functions. Finally
\begin{equation}
F_{\eta} = \eta \left( h - \frac{C_{t} \ln x}{2\tau C - 2 \chi_{0}} + C_{t} w  \right) , \ x:= \eta e^{2 \tau f}
\end{equation}
Because $F_{\eta \eta \eta \eta} \sim C_{t}$ there is only one condition for both SD and ASD curvature being nonzero and it reads $C_{t} \ne 0$.

The second possibility involves $a=b=1$ so homothetic vector has the form
\begin{equation}
K_{1} = \partial_{w} + \partial_{t} - 2 \chi_{0} \left( \phi \partial_{\phi} + \eta \partial_{\eta} \right)
\end{equation}
what implies the solution 
\begin{eqnarray}
&& C=f_{z} , \ f=f(z) , \ z:=w-t  , \  (\tau C^{2} + C_{z})_{z} \ne 0
\\ \nonumber
&& F_{\eta} = \eta (g-C) , \ g=g(y), \ y:= x e^{2\chi_{0}t} , \ x:= \eta e^{2 \tau f}   , \ (yg)_{yyy} \ne 0 
\end{eqnarray}

\subsection{Two symmetries}

As a starting point we take the results from the subsection \ref{jedna_symetria_NexNe_plusminus} written for $\chi_{0}=0$.

\subsubsection{Case $B=1$, $K_{1} = \partial_{w}$}

One arrives at the following form of the second, proper homothetic or Killing vector
\begin{equation}
 K_{2} = \left\{
   \begin{array}{ll}
  w \partial_{w} + \chi_{0} t \partial_{t} - \chi_{0} \phi \partial_{\phi} - \eta \partial_{\eta} & \textrm{for } \chi_{0} \ne 0
  \\
  w \partial_{w} + b_{0} \partial_{t} - \eta \partial_{\eta} & \textrm{for } \chi_{0} = 0
  \end{array}   \right.
\end{equation}
and the function $m$ reads
\begin{equation}
 m(t) = \left\{
   \begin{array}{ll}
  m_{0} t^{- \frac{2}{\chi_{0}}} & \textrm{dla } \chi_{0} \ne 0
  \\
  m_{0} e^{- \frac{2t}{b_{0}}} & \textrm{dla } \chi_{0} = 0
  \end{array}   \right.
\end{equation}
The solution for $F_{\eta}$ is exactly the same as (\ref{funkcja_F_po_eta_dla_NexNe_plusminus}) but with $g=g_{0} = \textrm{const}$.

\subsubsection{Case $B=1$, $K_{1} = \partial_{w} + \partial_{t}$}

Consider first the case with the second vector being the proper homothetic one, $\chi_{0} \ne 0$. It can be brought to the form
\begin{equation}
K_{2} = \chi_{0} \left( w \partial_{w} +  t \partial_{t} - \phi \partial_{\phi} - \eta \partial_{\eta} \right)
\end{equation}
With this second symmetry assumed, the field equation (\ref{zredukowane_rownanie_pola_dla_dla_NexNe_plusminus}) can be solved. Functions $m(x)$ and $F_{\eta} (\eta,x)$ are
\begin{eqnarray}
\label{pomocnicze1_dwie_symetrie_NexNe_plusminus}
&& m(x) = m_{0} x^{-2} , \ x:=w-t , \ m_{0} \ne 0
\\ \nonumber
&& F_{\eta} =   - \frac{2}{\tau} m_{0} x^{-2}  e^{-h} \int \frac{e^{h} dy}{2y^{2} - \tau^{-1} y +2m_{0}}  , \ h(y) :=  \int \frac{(2 \tau^{-1} - 4y)dy}{2y^{2} - \tau^{-1} y +2m_{0}}, \ y:=x \eta 
\\ \nonumber
&& \left(  e^{-h} \int \frac{e^{h} dy}{2y^{2} - \tau^{-1} y +2m_{0}}  \right)_{yyy} \ne 0
\end{eqnarray}
Solution for $F_{\eta}$ is nasty, but it can be written in the simpler form. To do this we use the ambiguity in the constant $\tau$ and we put $4m_{0}^{\frac{1}{2}} \tau =- 1$. The integrals in (\ref{pomocnicze1_dwie_symetrie_NexNe_plusminus}) can be calculated and we arrive at the formulas 
\begin{equation}
F_{\eta} = -2m_{0}x^{-2} s^{-2} (s^{2} - 2s +2 +h_{0} e^{-s}) , \ s:= \frac{2m_{0}^{\frac{1}{2}}}{x \eta +m_{0}^{\frac{1}{2}}} , \ h_{0} \ne 0
\end{equation}

However, if the second symmetry is given by a Killing vector, we find its form as
\begin{equation}
K_{2} = e^{a_{0} w} \left( \partial_{w} + \left( \frac{a_{0}^{2}}{4 \tau} - a_{0}\eta \right) \partial_{\eta} \right)
\end{equation}
where $a_{0} \ne 0$ is a constant. Functions $m$ and $F_{\eta}$ read
\begin{eqnarray}
&& m(x) = m_{0} e^{-2a_{0} x} - \frac{a_{0}^{2}}{16 \tau^{2}} , \ x := w-t , \ m_{0} \ne 0 , \  a_{0} \ne 0
\\ \nonumber
&& F_{\eta} = -e^{-a_{0}x} \left( \frac{a_{0}y}{2 \tau} + \frac{a_{0} (m_{0} + y^{2})}{2 \tau m_{0}^{\frac{1}{2}}} \left( \textrm{arctg} \, \left( \frac{y}{m_{0}^{\frac{1}{2}}} \right) + h_{0}  \right) \right)  , \ y:=e^{a_{0}x} \left( \eta - \frac{a_{0}}{4 \tau} \right)
\end{eqnarray}

\subsubsection{Case $B=0$, $K_{1} = \partial_{w} $}

The second homothetic vector can be brought to the form
\begin{equation}
K_{2} = w \partial_{w} + t \partial_{t} + (1-2\chi_{0}) \left( \phi \partial_{\phi} + \eta \partial_{\eta} \right)
\end{equation}
Now the solution is
\begin{equation}
F_{\eta} = \frac{\eta}{2\tau t} \left( h_{0} + \ln ( \eta t^{2\chi_{0}-1})  \right) , \ C(t) = C_{0} t^{-1} , \ C_{0} \ne 0
\end{equation}

\subsubsection{Case $B=0$, $K_{1} = \partial_{w} + \partial_{t} $}

In this case the second homothetic vector has the form
\begin{equation}
K_{2} = w \partial_{w} + t \partial_{t} + (1-2 \chi_{0}) \left( \phi\partial_{\phi} + \eta \partial_{\eta} \right) - \tau \epsilon \partial_{\eta}
\end{equation}
Then we get
\begin{equation}
 F_{\eta} = \eta (g-C) , \ g=g(x), \  x:= \eta z^{2 \tau C_{0}}  , \ C=C_{0} z^{-1} , \ z:=w-t , \ C_{0} \ne \left\{ 0, \frac{1}{\tau} \right\}
\end{equation}
Functions $\epsilon$ and $g$ read
\begin{equation}
\epsilon (z) = \left\{
   \begin{array}{ll}
  0 & \textrm{dla } \chi_{0} \ne \{ \tau C_{0}, \tau C_{0} + \frac{1}{2} , \tau C_{0} +1  \}
  \\
  \epsilon_{0} z^{- 2t C_{0}} & \textrm{dla } \chi_{0} = \tau C_{0} + \frac{1}{2} , \epsilon_{0} \ne \{ 0, \frac{1}{\tau}, \frac{1}{2\tau} \}
  \end{array}   \right.
\end{equation}

\begin{equation}
g(x) = \left\{
   \begin{array}{ll}
  g_{0} x^{\frac{1}{2\chi_{0}-2\tau C_{0} -1}} & \textrm{dla } \chi_{0} \ne \{ \tau C_{0}, \tau C_{0} + \frac{1}{2} , \tau C_{0} +1  \} , \ g_{0} \ne 0
  \\
  g_{0} x^{\frac{1-\tau \epsilon_{0}}{\tau \epsilon_{0}}}  & \textrm{dla } \chi_{0} = \tau C_{0} + \frac{1}{2} , \epsilon_{0} \ne \{ 0, \frac{1}{\tau}, \frac{1}{2\tau} \} , \ g_{0} \ne 0
  \end{array}   \right.
\end{equation}


\setcounter{equation}{0}
\section{Type $\{ [\textrm{N}]^{e} \otimes [\textrm{N}]^{n},[++] \}$}
\label{Sekcja_klasa_III_szczegolna}

The spaces considered in this section provide us with the second example of the spaces which do not posses real Lorentzian slices, but only neutral slices. The congruence of the null geodesics is twisting. General form of the key function (\ref{funkcja_kluczowa_dla_NexNn}) generates the metric
\begin{eqnarray}
\label{metryka_typu_NexNn_bez_uproszczen}
ds^{2} &=& 2\phi^{-2} \big\{ \tau^{-1} (d \eta  d w - d \phi  dt) -    \phi (\phi T_{xx}+A )\,   
dt^{2} \ \ \ \ \ \ 
\\ \nonumber
&& \ \ \ \ \ \ \ \ \ \ + ( 2 \phi \eta T_{xx} -2 \phi^{2} T_{x} +2A \eta) \, dw dt
  \\ \nonumber
&& \ \ \ \ \ \ \ \ \ \ +  ( 2\phi \eta T_{x} - \eta^{2} T_{xx} - \tau^{-1}C  \phi +2B \eta ) \, dw^{2} \big\} \ \ \ \ \ \ 
\end{eqnarray}
Obviously, the function $D$ does not enter into the metric and it can be gauged away, what we treat as done in the next sections.

\subsection{Generic case; no symmetries are assumed}

From (\ref{NexNn_e1}) it follows that $A=r_{t}$ and $B=r_{w}$, where $r=r(w,t)$ is an arbitrary function. Careful analysis of the formula (\ref{transformacja_funkcji_kluczowej}) leads to the transformation rule for $r$
\begin{equation}
\label{transformacje_na_r_typ_NexNn_plusplus}
r' = r + \frac{1}{2 \tau} \ln ( \lambda w'_{w})
\end{equation}
Hence, $r$ can be gauged away what implies $A=B=0$. The key function reads now
\begin{equation}
\label{funkcja_kluczowa_dla_NexNn_pouproszczeniach}
W(\phi, \eta, w,t) = \phi^{3} \, T(x,w,t)  - \frac{C (w,t)}{2\tau} \phi^{2} 
\end{equation}
and the field equation (\ref{NexNn_e3}) reduces to the form
\begin{equation}
\label{rownanie_pola_typ_NexNn_bez_symetrii}
 C \, T_{xx}  + T_{xw}-3T_{t}+xT_{xt}=0
\end{equation}
Gathering, the metric of the complex Einstein spaces of the type $\{ [\textrm{N}]^{e} \otimes [\textrm{N}]^{n},[++] \}$ can be always brought to the form
\begin{eqnarray}
\label{metryka_typu_NexNn}
ds^{2} &=& 2\phi^{-2} \big\{ \tau^{-1} (d \eta  d w - d \phi  dt) -    \phi^{2}  T_{xx} \,   
dt^{2} \ \ \ \ \ \ 
\\ \nonumber
&& \ \ \ \ \ \ \ \ \ \ + ( 2 \phi \eta T_{xx} -2 \phi^{2} T_{x} ) \, dw dt
  +  ( 2\phi \eta T_{x} - \eta^{2} T_{xx} - \tau^{-1}C  \phi  ) \, dw^{2} \big\} \ \ \ \ \ \ 
\end{eqnarray}
where function $T(x,w,t)$ satisfies Eq. (\ref{rownanie_pola_typ_NexNn_bez_symetrii}) and $C=C(w,t)$ is an arbitrary function such that $C_{tt} \ne 0$. Structural function $\gamma = C_{t}$.

\textbf{Remark}. Interesting fact is, that the vacuum Einstein equations in the spaces of the type $\{ [\textrm{N}]^{e} \otimes [\textrm{N}]^{e},[++] \}$ give an overdetermined system of three equations for two functions of three variables (see, e.g., \cite{Chudecki_Przanowski}). The lack of expansion of the congruence of ASD null strings allows to reduce vacuum Einstein field equations to the single equation (\ref{rownanie_pola_typ_NexNn_bez_symetrii}) for one function $T$ of three variables. It proves, that existence of the nonexpanding congruence of null strings is a very strong geometrical constraint.

\subsection{One symmetry}
\label{subsection_NexNn_plusplus_onesymmetry}

In this section we equip the space of the type $\{ [\textrm{N}]^{e} \otimes [\textrm{N}]^{n},[++] \}$ with additional symmetry given by the homothetic vector. Putting the general form of the key function (\ref{funkcja_kluczowa_dla_NexNn}) into the master equation (\ref{master_equation}) one finds $\epsilon=\beta=0$ and the equations which relate the functions $A,B,C$ and $T$ with functions $a,b$ and $\alpha$
\begin{subequations}
\label{ogolne_rownania_wynikajace_z_master_typ_[N]ex[N]n}
\begin{eqnarray}
\label{ogolne_rownania_wynikajace_z_master_typ_[N]ex[N]n_1}
&& aT_{w} + bT_{t} + ((b_{t}-a_{w})x +b_{w}) T_{x} + (2a_{w} - 2 \chi_{0}) T - \alpha = 0
\\ 
\label{ogolne_rownania_wynikajace_z_master_typ_[N]ex[N]n_2}
&& aA_{w} + bA_{t} + b_{t} A + \frac{1}{\tau} b_{tt} = 0
\\ 
\label{ogolne_rownania_wynikajace_z_master_typ_[N]ex[N]n_3}
&& aB_{w} + b B_{t} + a_{w} B +b_{w}A - \frac{1}{2 \tau} (a_{ww}- 2b_{tw})=0
\\
\label{ogolne_rownania_wynikajace_z_master_typ_[N]ex[N]n_4}
&& aC_{w} + b C_{t} - 2\tau B b_{w}+ (2a_{w}-b_{t}) C - b_{ww} = 0
\end{eqnarray}
\end{subequations}
[Note, that we do not use here simplifications which were essential in the case with no symmetry $(A=B=0)$. The reason why we keep $A$ and $B$ nonzero is as follows. The gauge freedom which is offered by the function $\lambda (w,t)$ plays a very important role in simplification of the form of homothetic vector. It is impossible to use $\lambda$ to make $A=B=0$ and to simplify the homothetic vector simultaneously. It appears, that better solution is to utilitize $\lambda$ to simplify the homothetic vector.] 

Detailed analysis gives us the form of the homothetic vector as
\begin{equation}
 K_{1} = \partial_{w}  -2 \chi_{0}  \left( \phi \partial_{\phi} +  \eta  \partial_{\eta} \right)
\end{equation}
Eqs. (\ref{ogolne_rownania_wynikajace_z_master_typ_[N]ex[N]n_2}-\ref{ogolne_rownania_wynikajace_z_master_typ_[N]ex[N]n_4}) together with Eq. (\ref{NexNn_e1}) imply $A=A(t)$, $B=B_{0} = \textrm{const}$, $C=C(t)$. Eq. (\ref{ogolne_rownania_wynikajace_z_master_typ_[N]ex[N]n_1}) gives $T(x,w,t) = e^{2\chi_{0} w} H(x,t) + \tilde{H}(w,t)$. Analysis of the field equation (\ref{NexNn_e3}) shows that $\tilde{H}(w,t) = e^{2\chi_{0} w} (f_{1}(w) + f_{2} (t))$, but $f_{1}(w)$ can be gauged away with the help of the gauge function $L(w)$ and $f_{2} (t)$ can be absorbed into $H(x,t)$. Finally, the key function takes the form
\begin{equation}
W(\phi, \eta, w,t) = \phi^{3} e^{2 \chi_{0} w} H(x,t) +\frac{A(t)}{2} \eta^{2} + B_{0} \eta \phi  - \frac{C(t)}{2 \tau} \phi^{2} , \ x:= \frac{\eta}{\phi}
\end{equation}
The gauge freedom still available enables us to put $A=0$ without any loss of generality. The last equation that remains to be solved is the Eq. (\ref{NexNn_e3}), which reads now
\begin{equation}
\label{ostateczne_equation_NexNn_plus_plus}
(-2 \tau B_{0} x +  C) H_{xx} + (2 \tau B_{0} + 2 \chi_{0}) H_{x} + xH_{xt} - 3 H_{t} = 0
\end{equation}
Structural function $\gamma(t)$ takes the form
\begin{equation}
\frac{\gamma}{\tau^{2}} = B_{0}^{2} + \frac{1}{\tau^{2}} C_{t} , \ C_{tt} \ne 0
\end{equation}

Gathering, the metric of the vacuum space of type $\{ [\textrm{N}]^{e} \otimes [\textrm{N}]^{n},[++] \}$ with one symmetry given by the homothetic vector $ K_{1}$ can be brought to the form
\begin{eqnarray}
\label{metryka_typu_NexNn_jednasymetria}
ds^{2} &=& 2\phi^{-2} \big\{ \tau^{-1} (d \eta  d w - d \phi  dt) -    \phi^{2} e^{2\chi_{0}w} H_{xx} \,   
dt^{2} 
\\ \nonumber
&& \ \ \ \ \ \ \ \ \ \ 
+ 2 e^{2\chi_{0}w} \phi^{2} (x H_{xx} -H_{x} ) \, dw dt
\\ \nonumber
&& \ \ \ \ \ \ \ \ \ \ 
  +  ( 2 B_{0} \eta  - \tau^{-1} C \phi - e^{2\chi_{0}w} \eta \phi (xH_{xx} - 2 H_{x} )) \, dw^{2} \big\} \ \ \ \ \ \ 
\end{eqnarray}
where $C(t)$ is an arbitrary function such that $C_{tt} \ne 0$ and $H(x,t)$ satisfies the equation (\ref{ostateczne_equation_NexNn_plus_plus}).

\subsection{Two symmetries}

The detailed discussion concerning type $\{ [\textrm{N}]^{e} \otimes [\textrm{N}]^{n},[++] \}$ with two homothetic symmetries has been presented in \cite{Chudecki_Przanowski}. Hence, we collect here only brief summary of the results from \cite{Chudecki_Przanowski}. As a starting point we take the results from the subsection \ref{subsection_NexNn_plusplus_onesymmetry} written for the $\chi_{0}=0$ (the first Killing vector is $K_{1}=\partial_{w}$). Moreover, we do not gauge away function $A$. The key function reads now
\begin{equation}
W(\phi, \eta, w,t) = \phi^{3} H(x,t) +\frac{1}{2} A(t) \eta^{2} + B_{0} \eta \phi  - \frac{C(t)}{2 \tau} \phi^{2} , \ x:= \frac{\eta}{\phi}
\end{equation}
the structural function is
\begin{equation}
\frac{\gamma}{\tau^{2}} = B_{0}^{2} + \frac{AC}{\tau} + \frac{C_{t}}{\tau^{2}}  
\end{equation}
and $H(x,t)$ satisfies the equation
\begin{equation}
\label{rownanie_pomocccniczne}
(-\tau A x^{2} -2 \tau B_{0} x +  C) H_{xx} + (2\tau Ax + 2 \tau B_{0} ) H_{x} + xH_{xt} - 3 H_{t} = 0
\end{equation}

Further steps consist of analysis of Eqs. (\ref{ogolne_rownania_wynikajace_z_master_typ_[N]ex[N]n}) written for the second homothetic vector. The analysis is long and tedious but finally we arrive at the following form of $K_{2}$
\begin{equation}
K_{2} = w \partial_{w} + t \partial_{t} + (1-2\chi_{0}) \left( \phi \partial_{\phi} + \eta \partial_{\eta} \right)
\end{equation}
and we get 
\begin{equation}
A(t) = A_{0} t^{-1}, \ B_{0}=0 , \ C(t) = C_{0} t^{-1}
\end{equation}
The key function reads
\begin{equation}
W(\phi, \eta, w,t ) = \phi^{3} ( t^{2\chi_{0}-2} u(x) + \tilde{H} (t) ) + \frac{A_{0}}{2} \frac{\eta^{2}}{t} - \frac{C_{0}}{2 \tau} \frac{\phi^{2}}{t}
\end{equation}
Function $\tilde{H} (t)$ satisfies the relation $3 \tilde{H}_{t} = - \tau H_{0} t^{2\chi_{0}-3}$ where $H_{0}$ is some constant. However this equation is not important at all, because $\tilde{H} (t)$ does not enter into the metric. Structural function $\gamma$ reads
\begin{equation}
\gamma (t) = (\tau A_{0} - 1) C_{0} t^{-2} , \ C_{0} \ne 0, \  \tau A_{0} - 1 \ne 0
\end{equation}
The field equation (\ref{rownanie_pomocccniczne}) reduces to the equation for $u(x)$ 
\begin{equation}
\label{ostateczne_rownanie_typ_NexNn_plusplus_dwie_symetrie}
(A_{0} x^{2} - \tau^{-1} C_{0}) \frac{d^{2}u}{dx^{2}} - 2 \tau^{-1} (\tau A_{0} +\chi_{0} -1) x \frac{du}{dx} + 6 \tau^{-1} (\chi_{0}-1) u + H_{0} = 0
\end{equation}
The metric takes the form
\begin{eqnarray}
\label{metryka_typ_NexNn_plusplus_dwie_symetrie}
 ds^{2} \! &=& \! 2 \phi^{-2} \left\{ \frac{1}{\tau} (d \eta dw - d\phi dt) - \phi \left( \frac{A_{0}}{t} + \phi t^{2 \chi_{0}-2} \frac{d^{2} u}{dx^{2}} \right) dt^{2} \right.
\\ \nonumber
&& \ \ \ \ \ \ 
+\left( 2 \phi^{2} t^{2 \chi_{0}-2} \left( x \frac{d^{2} u}{dx^{2}} - \frac{du}{dx} \right) + \frac{2A_{0}}{t} \phi x  \right) dw dt
\\ \nonumber  
&& \ \ \ \ \ \ 
+\left. \left(  \phi^{2} t^{2 \chi_{0}-2} x \left( 2 \frac{du}{dx} -x \frac{d^{2} u}{dx^{2}}  \right) - \frac{C_{0}}{\tau t} \phi   \right) dw^{2} \right\}
\end{eqnarray}
The metric (\ref{metryka_typ_NexNn_plusplus_dwie_symetrie}) does not depend on $u$, but on $du / dx$. Denoting $U(x) := du / dx $ we arrive at the equation 
\begin{equation}
\label{ostateczne_rownanie_typ_NexNn_plusplus_dwie_symetrie_2}
(A_{0} x^{2} - \tau^{-1} C_{0}) \frac{d^{2}U}{dx^{2}} - \frac{2}{\tau} (\chi_{0} -1) x \frac{dU}{dx} + \left( \frac{4}{\tau} (\chi_{0}-1) - 2 A_{0} \right) U=0
\end{equation}
Both equations (\ref{ostateczne_rownanie_typ_NexNn_plusplus_dwie_symetrie}) and (\ref{ostateczne_rownanie_typ_NexNn_plusplus_dwie_symetrie_2}) have been solved in \cite{Chudecki_Przanowski}.

\setcounter{equation}{0}
\section{Concluding remarks}

In the present paper we have investigated the complex and real vacuum type $[\textrm{N}] \otimes [\textrm{N}]$ spaces. Well known classification of such spaces uses the properties of the congruences of SD and ASD null strings. In our approach additional criterion has been used: we consider the properties of the intersections of these congruences. Such approach leads to six distinct types of the $[\textrm{N}] \otimes [\textrm{N}]$ spaces (see Table \ref{Tabela_typy_N}). The most general one has been discussed in our previous paper \cite{Chudecki_Przanowski}. Remaining five types are analyzed from the point of view of the different signatures of the metric. Also, homothetic and Killing symmetries of these types are considered.

Although vacuum type [N] spaces have been considered many times by many authors, we were able to obtain a few interesting and new results:
\begin{enumerate}
\item Three examples of the Lorentzian slices of the complex metrics are found. The first one (the metric (\ref{metryka_NexNe_minus_minus_ciecie_Lorent})) leads to the Kundt class. The second one (the metric (\ref{metryka_i_rownanie_klasa_Rob_Traut_3})) and the third one (the metric (\ref{general_form_metric_NexNe_plusminus_step2})) lead to the Robinson - Trautman solution.
\item Some new interpretation of the difference between pp-waves class and Kundt class as considered on the level of  complexification of these spaces is given. It appears, that congruences of SD and ASD null strings of these classes have different properties (for pp-wave metrics both these congruences are nonexpanding, for the Kundt class both are expanding).
\item Two types of spaces which are equipped with the congruences of SD and ASD null strings of the different properties ($\{ [\textrm{N}]^{e} \otimes [\textrm{N}]^{n}, [--] \}$,$\{ [\textrm{N}]^{e} \otimes [\textrm{N}]^{n}, [++] \}$) are investigated in all details. Such types posses only real neutral slice. Especially interesting is the type $\{ [\textrm{N}]^{e} \otimes [\textrm{N}]^{n}, [++] \}$ which admits the congruence of null geodesics with nonzero twist.
\end{enumerate}

The further investigations could be focused on the following issues:
\begin{enumerate}
\item The similarity between the equation which describes the Hauser solution and the equation (\ref{ostateczne_rownanie_typ_NexNn_plusplus_dwie_symetrie_2}) is obvious. It suggests that similarities between type $\{ [\textrm{N}]^{e} \otimes [\textrm{N}]^{e}, [++] \}$ and $\{ [\textrm{N}]^{e} \otimes [\textrm{N}]^{n}, [++] \}$ can appear also in the cases with only one symmetry and without any symmetries. The importance of the vacuum type $\{ [\textrm{N}]^{e} \otimes [\textrm{N}]^{e}, [++] \}$ in models of gravitational waves justifies further investigations of similarities between these two types. Particularly the natural question arises: if the vacuum Einstein field equations for the type $\{ [\textrm{N}]^{e} \otimes [\textrm{N}]^{e}, [++] \}$ with no symmetries can be reduced to a single equation?
\item No explicit examples of the spaces of the type $\{ [\textrm{N}]^{e} \otimes [\textrm{N}]^{n}, [++] \}$ with no symmetries and with one symmetry have been found so far. Eqs. (\ref{rownanie_pola_typ_NexNn_bez_symetrii}) and (\ref{ostateczne_equation_NexNn_plus_plus}) are so interesting that they deserve further investigations.
\item All considerations could be generalized to the case with nonzero cosmological constant.
\end{enumerate}


\end{document}